\documentclass {aastex62}
\graphicspath{{./}{figures/}}

\submitjournal{PSJ}

\shorttitle{Chondrules}
\begin{document}

\title{The Macroporosity of Rubble Pile Asteroid Ryugu and Implications for the Origin of Chondrules}

\correspondingauthor{William Herbst}
\email{wherbst@wesleyan.edu}

\author [0000-0001-7624-3322]{William Herbst}
\affil{Department of Astronomy,
Wesleyan University,
Middletown, CT, 06457, USA}

\author {James P. Greenwood}
\affil{Department of Earth \& Environmental Sciences,
Wesleyan University,
Middletown, CT, 06457, USA}

\author {Teng Ee Yap}
\affil{Department of Physics and Astronomy,
Colgate University,
Hamilton, NY, 13346 USA}



\begin{abstract}

We use the known surface boulder-size distribution of the C-type rubble pile asteroid Ryugu (NEA 162173) to determine its macroporosity, assuming it is a homogeneous granular aggregate. We show that the volume-frequency distribution of its boulders, cobbles and pebbles, is well represented by a lognormal function with $\sigma = 2.4 \pm 0.1$ and  $\mu = 0.2 \pm 0.05$. Application of linear-mixture packing theory yields a value for the macroporosity of $\phi = 0.14 \pm 0.04$. Given its low bulk density of 1.19 gm cm$^{-3}$, this implies an average density for Ryugu's rocks of $1.38 \pm 0.07$ gm cm$^{-3}$ throughout its volume, consistent with a recent determination for surface boulders based on their thermal properties. This supports the spectrum-based argument that IDP's may be the best analog material available on Earth and suggests that high-density, well-lithified objects such as chondrules and chondrule-bearing chondrites may be rare on Ryugu. Implications of this result for the origin of chondrules, a long-standing problem in cosmochemistry, are discussed. We propose that chondrules and most chondrites formed together in rare lithification events, which occurred during the accretion of chondritic envelopes to large, differentiated planetesimals at a time when they were still hot from $^{26}$Al decay. 

\end{abstract}



\section{Introduction} \label{sec:intro}

Asteroids and meteorites can be associated with one another by their reflectance spectra \citep{GB20}. However, while overall composition may be similar, the bulk density ($\rho_{bulk}$) of small asteroids is often significantly less than the density of their analog meteorites \citep{B02}. For example, the meteorite analogs of 162173 Ryugu and 101955 Bennu, are carbonaceous chondrites (CC), whose densities, with the exception of the rare CI class, range from 2.2 to 5.3 gm cm$^{-3}$ \citep{M11}, well above the asteroids' bulk densities of $1.19 \pm 0.02$ and $1.19 \pm 0.013$ gm cm$^{-3}$, respectively \citep{Wa19,L19}. Similarly, the S-type asteroid 25143 Itokawa has $\rho_{bulk} = 1.95 \pm 0.14$ gm cm$^{-3}$ \citep{A06}, well below its meteorite analog, LL ordinary chondrites (OC), which have densities in the range 3.15 - 3.4 gm cm$^{-3}$ \citep{C06}. The standard explanation \citep{B02} has been that the asteroids are `rubble piles', i.e. loosely packed piles of gravitationally reassembled pieces of a pre-existing object that suffered a major impact. An assumption behind this interpretation is that asteroidal rocks can be assigned the same density as their associated meteorites, which were, of course, collected on the Earth's surface. 

This assumption is challenged by recent results from the ongoing sample return missions Hayabusa2 \citep{Wa19} and OSIRIS-REx \citep{L17}, which visited Ryugu and Bennu, respectively. Based on its thermal properties one may infer that a Ryugu boulder studied by \citet{G19} and \citet{H20} has a porosity of between 0.30 and 0.52. This translates to a boulder density of between 1.2 and 1.7 gm cm$^{-3}$ if we assume a grain density appropriate to the closest spectral match to Ryugu, CI meteorites \citep{K19}. Global images of Ryugu indicate that such  high porosity boulders are very much the rule, not the exception on its surface \citep{O20}. \citet{G20} found an average density of around 1.4 gm cm$^{-3}$ for Ryugu rocks in their study of its packing density. During the sample retrievals it has become quite clear that the tensile strength of the surface rocks is very low compared with most, if not all, meteorites. \citet{G19} have shown that the Ryugu boulder they studied probably does not have the tensile strength to survive passage through the Earth's atmosphere. It may be, therefore, that these asteroids are not composed of material already familiar to us from meteorite collections, even though they have similar reflectance spectra. In particular, meteorites appear to have higher densities and greater tensile strength than common rocks on these asteroids. If confirmed, this may have implications for the densest and most abundant component of such meteorites, their chondrules. 

Chondrules are dense, mm-scale, igneous spherules that comprise 20-80\% of the volume of nearly all of their eponymous meteorite type, the chondrites \citep{S04, CJ16}. Since $\sim$90\% of all meteorites in our collections are chondrites, there is a common opinion that chondrules are abundant in primitive asteroids and in the early Solar System \citep{K18}. They are often referred to in text books as the `building blocks' of terrestrial planets and could play an important role in that process \citep{J15}. It is disappointing, therefore, that despite decades of concerted effort and an abundance of fine detail, there remains no consensus among cosmochemists on how they formed \citep{C05, D12, CJ16, R18}. Their textures indicate that they were born in one or more short blasts of heat, lasting only minutes to hours, but intense enough to melt pre-chondrule matter (requiring T$\geqslant$1600 K) without fully vaporizing it \citep{HR90,J18}. But the community is  divided on whether these heat blasts occurred in a `nebular' setting or a `planetary' setting \citep{CJ16,R18}. That is, was chondrule formation a widespread phenomenon within the protoplanetary disk that preceded asteroid and planet formation, or a much smaller scale phenomenon associated with the late stages of planetesimal formation? In the first case, one might expect primitive asteroids to be as full of chondrules as chondrites are. In the second, chondrules may be rare or absent on many such objects.       

\citet{S98} pointed out decades ago that an extremely rare class of chondrites on Earth, CI's, which contain no chondrules at all, may actually be the most common type of chondrite in space. Their low density and tensile strength make them much less likely to survive passage through the Earth's atmosphere and terrestial weathering, should they reach the Earth's surface. A space-based sample of primitive asteroidal material might, therefore, contain a much lower percentage of chondrules than the Earth-based sample represented by our museum collections. This point of view is buttressed by the fact that the flux of meteorites reaching the surface of the Earth is only $\sim 0.15$ metric tonnes per day, or less than 0.1-0.5\% of the 30 - 180 t of material impacting the Earth daily at the top of the atmosphere \citep{B96, Z06, D17}. Our meteorite collections clearly sample only a tiny fraction of the interplanetary material incident on the Earth, with most of it arriving in the form of interplanetary dust particles (IDP's) that are much smaller than chondrites or chondrules and could have an asteroidal or cometary origin.

Here we attempt to illuminate these issues by directly calculating the amount of void space in Ryugu's interior, based on what we know about granular piles on Earth, which then allows us to estimate the average density of its rocks. Space between constituent rocks in a pile is quantified by porosity ($\phi$), defined as 
\begin{equation}
\phi = \left(1 - {V_{rock} \over V_{bulk}}\right)
\end{equation}
where V$_{rock}$ is the volume occupied by the rocks (grains) themselves and V$_{bulk}$ is the volume occupied by the pile (asteroid). In planetary science this is generally referred to as the `macroporosity' of the pile to distinguish it from the porosity of the rocks themselves, known as `microporosity'.  Assuming that the constituent rocks have a single density ($\rho_{rock}$), and that the pore spaces are empty, equation 1 may be rewritten in terms of bulk and rock densities, as 
\begin{equation}
\phi = \left(1 - {\rho_{bulk} \over \rho_{rock}}\right).
\end{equation}
Application of this equation yields a porosity of about 0.4 for Itokawa \citep{A06} and 0.48 - 0.62 for Ryugu and Bennu if they are composed of rocks with the same density as LL OC's and common, chondrule-bearing CC's, respectively. If Ryugu and/or Bennu are composed of non-chondrule bearing CI material then their inferred macroporosity is much lower ($\sim0.24$), since CI density, based on a single measurement, is only 1.57 gm/cm$^3$ \citep{M11}. \citet{G20} recently reported a value of $\phi = 0.16 \pm 0.03$ for Ryugu based on a similar approach to the one adopted here. 

In this paper, we first calculate $\phi$ for Ryugu and then use Equation 2 to determine its average rock density, which can be compared with chondrules, meteorites and with the surface boulder measurement. We model the asteroid as a self-gravitating, homogeneous granular aggregate, whose grain size distribution can be inferred from surface boulder counts. We proceed by first deriving the boulder-size distribution that would apply to a representative volume within the surface layer of the asteroid, based on the surface boulder counts of \citet{M19}. We show that the distribution is well-represented by a lognormal function. This allows us to estimate the macroporosity of Ryugu by application of linear-mixture packing theory for a continuous frequency distribution \citep{YS91}. We then address the broader question of what the relationship between chondrites and their host asteroids may be. This impacts discussion of the long-standing and vexing problem of the origin of chondrules, which has implications for the structure and origin of asteroids and terrestrial planets, as we discuss. 

\section{The Porosity of a Granular Aggregate}

There is considerable evidence now that small, rubble pile asteroids such as Itokawa, Ryugu and Bennu are `granular aggregates', composed entirely of mobile, unconsolidated material in the form of boulders, cobbles, pebbles and perhaps finer grained material, held together by self-gravity and containing no separate, monolithic core  \citep{A06,S13, S14, J20, P20}.  Granular aggregates on Earth are common in river beds and beaches and have been studied extensively in a wide variety of contexts. They do not fit within the usual solid/liquid/gas classification of states of matter but exhibit a range of phenomena unique to themselves, including granular flow or convection \citep{JN92}. For example, large boulders, even when denser than their surroundings, rise to the top of shaken piles instead of sinking to their bottoms, as they would in a liquid or gas, a phenomenon known as the `Brazil Nut Effect' \citep{R87, M14}. 

The principal factors affecting the porosity of a granular aggregate on Earth are well known. Of greatest importance are grain size distribution, grain shape and how the grains are packed \citep{YS91}. In terrestrial piles, liquid often fills the pores and that can affect porosity. In the case of very small grains (a few hundred microns or less), cohesive effects can play a role due to the relative importance of electrostatic forces between grains \citep{YZY00,Z11}. Here we assume that the pores in a granular aggregate asteroid are empty and that cohesive forces between them are negligible. We return to a discussion of the latter point after establishing the size of the grains controlling porosity on Ryugu. As a geometric entity, porosity is independent of absolute scale unless forces other than gravity are involved \citep{YS91,F11}. While the surface gravity of a small asteroid is only about $10^{-5}$ g, its porosity-controlling grains (rocks) are meter-scale, or 1000 times larger than the mm-scale grains composing typical terrestrial aggregates. There is no reason of which we are aware that terrestrial results on porosity would not be applicable to granular piles in space, despite these scale differences.

The porosity of a pile with a single grain size depends on the grain shape and how they are packed. Spheres are best studied and  experimental, analytic and computational techniques give consistent results. The theoretical lower limit for spheres of a single size is $$\phi = {1 - {\pi \over 3 \sqrt 2}} = 0.26,$$ a result that was conjectured by Kepler and proven by Carl Friedrich Gauss in 1831 \citep{H98}. When spheres are randomly poured into a container the porosity of the pile will initially vary between 0.37 and  0.445 and if the container is shaken sufficiently it will eventually achieve a condition known as random close packing (RCP) characterized by $\phi = 0.3634 \pm 0.0005$ \citep{SK69}. Non-spherical particles in RCP typically have a porosity of around 0.40 unless the particle shapes are extreme (e.g. needles or flat disks) \citep{YZ98}. 

In binary mixtures or piles with a range of sizes $\phi$ can be lowered substantially by two factors that come into play. First, small grains tend to filter down through the spaces between larger grains, filling them and, thereby, reducing porosity. Second, large grains embedded in a sea of smaller ones displace not only the grains themselves but also the pore space between them, again lowering the overall porosity of the pile. The first effect is commonly known as infiltration or sifting and the second as occupation \citep{YS91}. The magnitude of these effects depends on the relative sizes and abundances of the grain mixtures, but the direction is always the same -- mixtures only lower the porosity of a pile, sometimes dramatically. A well-chosen size distribution, such as the `Fuller distribution' can reduce porosity to nearly zero, a fact that has been known for more than a century and is exploited commercially in making cement \citep{F07}. 

\citet{WH30} showed that, for binary mixtures of spherical particles with r$_1$ $>$ r$_2$, $\phi$ decreases rapidly to a minimum value of 0.16 as the ratio r$_1$/r$_2$ increases. The minimum value is achieved when the larger granules make up about 2/3 of the total volume. In that case, they create a framework with $\sim$36\% void space where the small particles congregate. Adding additional components has the effect of lowering the minimum porosity limit even further. For example, \citet{WH30} find that a three component mixture with a ratio of granule radii in the proportion 50.5:8:1 has a minimum porosity of only 0.053 when mixed in a volume proportion of 66\% large, 25\% medium and 9\% small particles. Mixtures with nearly equal volumes yield a porosity of 0.22. It is important to note that not only the grain size distribution, but also the degree to which the particles are mixed, determines the overall porosity of a pile. In a completely layered structure, for example, neither sifting nor occupation occur so the porosity of the pile assumes the value for a uniform size distribution. Interestingly, shaking a pile with a distribution of grain sizes can actually raise its porosity, not lower it, because of the Brazil Nut Effect, which tends to separate large particles (at the top) from smaller ones (at the bottom),  at least in the laboratory within walled containers.

Ideally, one would like to know the grain size  and shape distribution as a function of depth within a pile to most accurately predict its macroporosity. For asteroids, such information will not be known for a long time, if ever. However, under the assumptions that the object is well mixed and homogeneous we can use surface boulder counts and shape determinations to estimate reasonable limits on $\phi$. Then, equation 2 can be used to infer the grain (rock) density, which can be compared with direct measurements of surface rocks from Ryugu brought back by the Hayabysa2 mission, when they become available. If there is consistency between surface measurements and the `whole body' average determined by this approach it would support the assumption of homogeneity. That is the approach adopted here, and it begins with a determination of the boulder size distribution applicable to the surface layer of the asteroid.

\section{The Boulder-Size Distribution within Ryugu's Surface Layer}

 \citet{M19} report rock counts by size on the surface of Ryugu in the range 0.02 to 140 m.  From the counts they derive the relative size-frequency distribution, R, which applies to a unit area on the surface of the asteroid (see their figures 2b and 5b). Since the asteroid does not have an independent solid surface on which the boulders lie, but is a granular aggregate, the areal counts do not directly reflect what the size distribution in a unit volume of the surface layer would be. In particular, larger boulders are over-represented in areal counts because they can be seen to greater depths within the surface layer than smaller boulders, cobbles or pebbles. 
 
 \begin{figure}[ht!]
\plotone{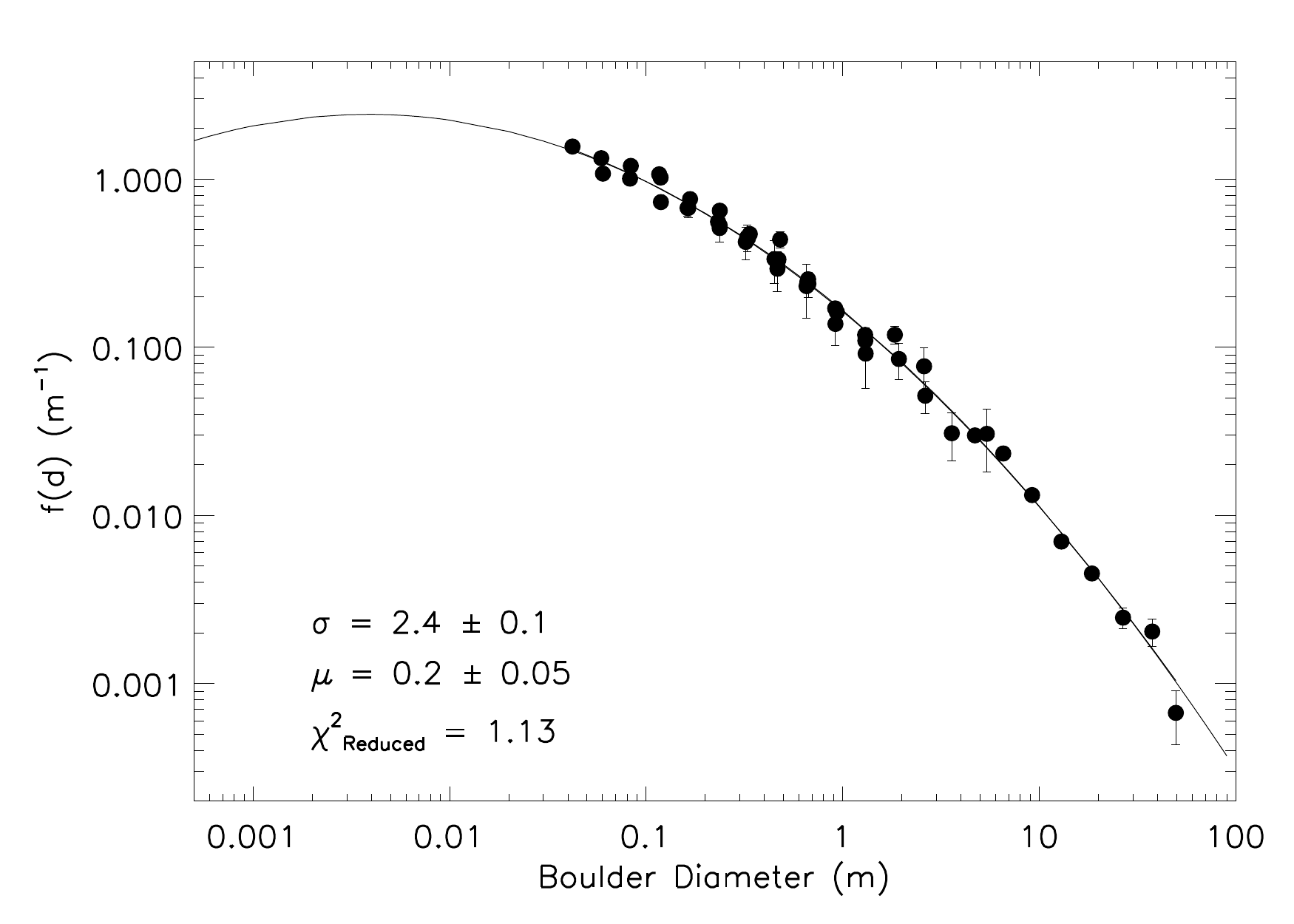}
\caption{The normalized volume frequency distribution for Ryugu boulders, derived from the areal counts of \citet{M19}, compared to a lognormal distribution with $\sigma = 2.4 \pm 0.1$ and $\mu = 0.2 \pm 0.05$. The fit is clearly quite good and characterized by a reduced $\chi^2 = 1.13$. Error bars based on counting statistics alone are shown. Where they do not appear, the error bars are smaller than the data points.}
\end{figure}

In Appendix A we derive the relationship between areal counts, quantified by R(d), and volume counts, quantified by $f(d)$, where $f(d)$ is the size-frequency distribution that applies to a unit volume of the asteroid's surface (see Equation A6). Since the boulder counts are done on images of vastly different scales, taken from a range of heights above the asteroid, they must be normalized before being compared with one another. In carrying out this process, we noticed the strong resemblance of $f(d)$, based on the boulder counts, to a lognormal distribution (see Fig. 5 of Appendix A). 

Since a lognormal distribution is the result of a natural process that involves many small, uncorrelated multiplicative/divisive events and since it appears as commonly in nature as its better-known analog the normal (Gaussian) distribution that results from a large number of uncorrelated addition/subtraction events \citep{L01}, this is not surprising. The surface of Ryugu is presumably frequently hit by meteoroids, which may fragment its rocks, leading to a representative volume sample of its surface having a lognormal boulder-size distribution no matter what their initial size distribution may have been. Or, the asteroid may have reassembled from a catastrophic distruction that endowed its remnants with a lognormal distribution. While the lognormal function is one of the most commonly employed representations in all fields of science, it has been particularly associated with fragmentation processes on a variety of scales, from rock crushing to the formation of stars in clusters and associations \citep{M79, B89}. It is worthwhile, therefore, to investigate exactly how closely $f(d)$ follows a lognormal distribution. 

\begin{figure}[ht!]
\plotone{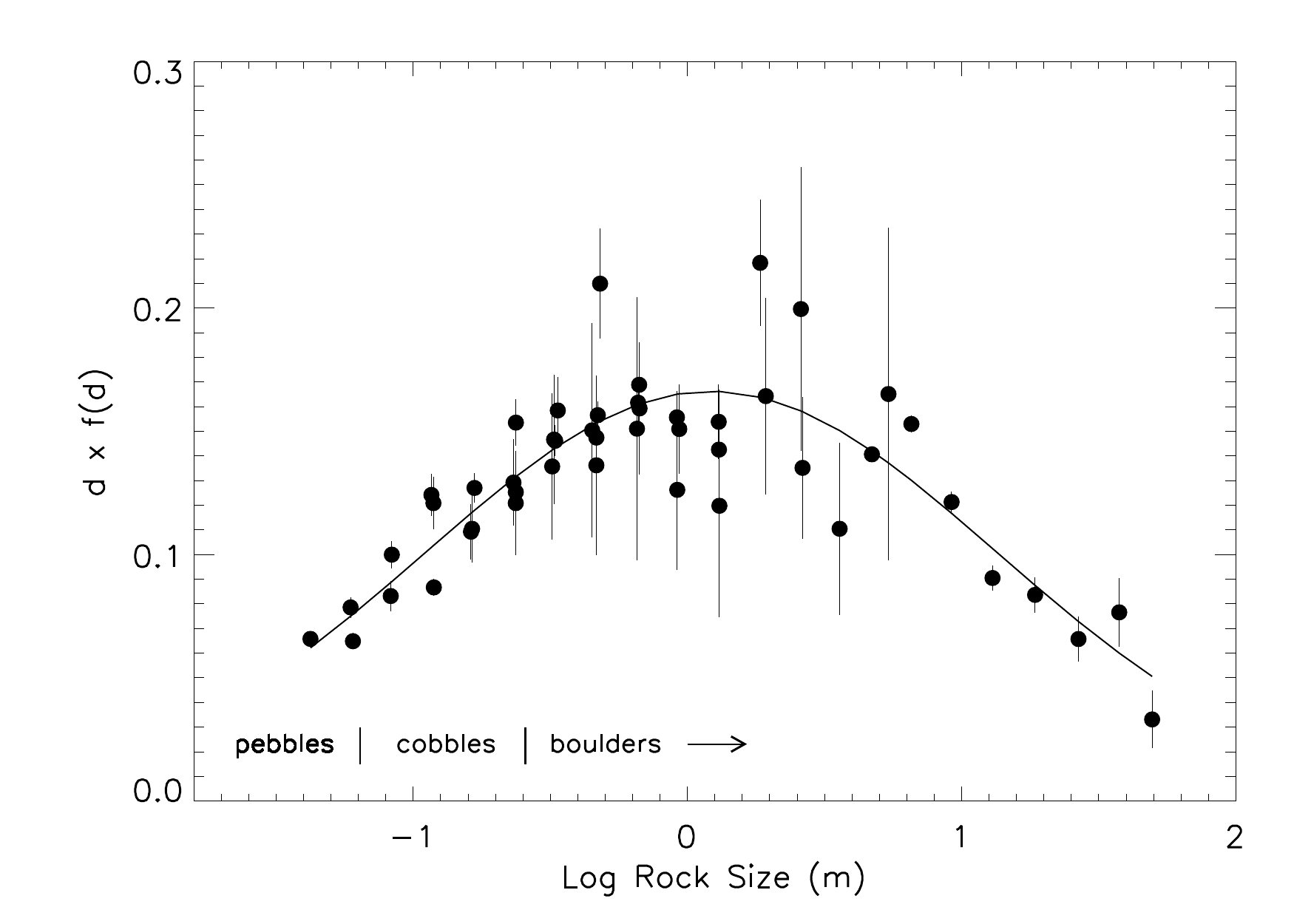}
\caption{$d \times f(d)$ versus log (d) showing the Gaussian form explicitly. The peak of the distribution is at boulders with d = exp(0.2) = 10$^{0.087}$ = 1.2 m. Error bars on the points reflect counting statistics alone and are, in some cases, smaller than the points themselves.} \end{figure}

The lognormal function may be written as:

\begin{equation}
{f(x)} = {1 \over { \sigma x \sqrt {2 \pi}}}\ {\rm exp} { \left( - {{(ln (x) - \mu)^2} \over {2 \sigma^2}}\right)}.
\end{equation}
It has two free parameters, $\sigma$ and $\mu$, and is normalized such that $\int_0^\infty f(x) dx = 1.$ In Figure 1 we compare our measurements of $f(d)$, renormalized to a common scale (see Appendix B), based on the surface boulder counts, to a best-fit lognormal distribution, determined by a non-linear Levenberg-Marquardt fitting technique. The data points are weighted by the number of boulders counted in each bin and error bars reflect only counting statistics. The fitting process returns the following values and error limits on the parameters: $\mu = 0.20 \pm 0.05$ and $\sigma = 2.4 \pm 0.1$. By eye, the lognormal distribution is clearly a good fit to the data over its full range, approaching four orders of magnitude in size. This impression is verified quantitatively by the reduced $\chi^2$ value that applies to the fit, $\chi^2 = 1.13$. The reduced chi-square is the chi-square per degree of freedom and a value of 1 is expected of a perfect fit to data whose error bars are accurately known. Since we have only included Poisson (counting) effects in the estimation of errors on the individual data bins here, and since there are only two free parameters ($\mu$ and $\sigma$) involved in the fit, this is about as good as one could expect. It would seemingly be hard to improve on the lognormal function as an appropriate representation of these data, especially given its physical underpinning.

In Figure 2 we display the same data and fit in a different manner that highlights its Gaussian nature and makes the quality of the fit more apparent. In this case, the ordinate ($d \times f(d)$) is unitless and is the fraction of the sample that would be contained within a bin centered on the specified boulder size. The distribution peaks at a value of exp(0.2) = 1.2 m, indicating that in a representative sample of the surface layer of Ryugu, boulders a little larger than 1 m are the most common contributors to the total volume or mass. The fact that its surface layer is so well represented as lognormal greatly simplifies the task of estimating the porosity of Ryugu under the assumption that it is an homogeneous object. The size scale of the boulders controlling the porosity ($\sim$1 m) also informs the discussion of the extent to which mobility issues and/or cohesion among grains may or may not be important to consider in estimating the macroporosity of Ryugu, a task to which we now turn.  

\section{The Macroporosity of Ryugu and Average Density of its Boulders}

With $f(d)$ established for the surface layer we can proceed to estimate the porosity of Ryugu under the assumption that it is an homogeneous object. \citet{F11} have evaluated a number of methods developed for terrestrial piles and found one, the `linear-mixture' packing theory of \citet{YS91}, to be the most reliable and we employ it here. It is a semi-empirical theory that can be applied to either discrete or continuous frequency distributions. The authors have provided a convenient representation of the predictions of their model for the lognormal distribution because of its widespread appearance as a descriptor applicable to terrestrial piles. As they show in their Figure 12, the macroporosity ($\phi$) of a pile with a lognormal grain-size distribution depends on only two parameters, the geometric standard deviation, $\sigma_G$, and the `initial porosity', $\phi_0$, of the grains. Note, in particular, that the macroporosity is independent of the median particle size, $\mu$, as long as the initial porosity is independent of grain size. 

\begin{figure}[ht!]
\plotone{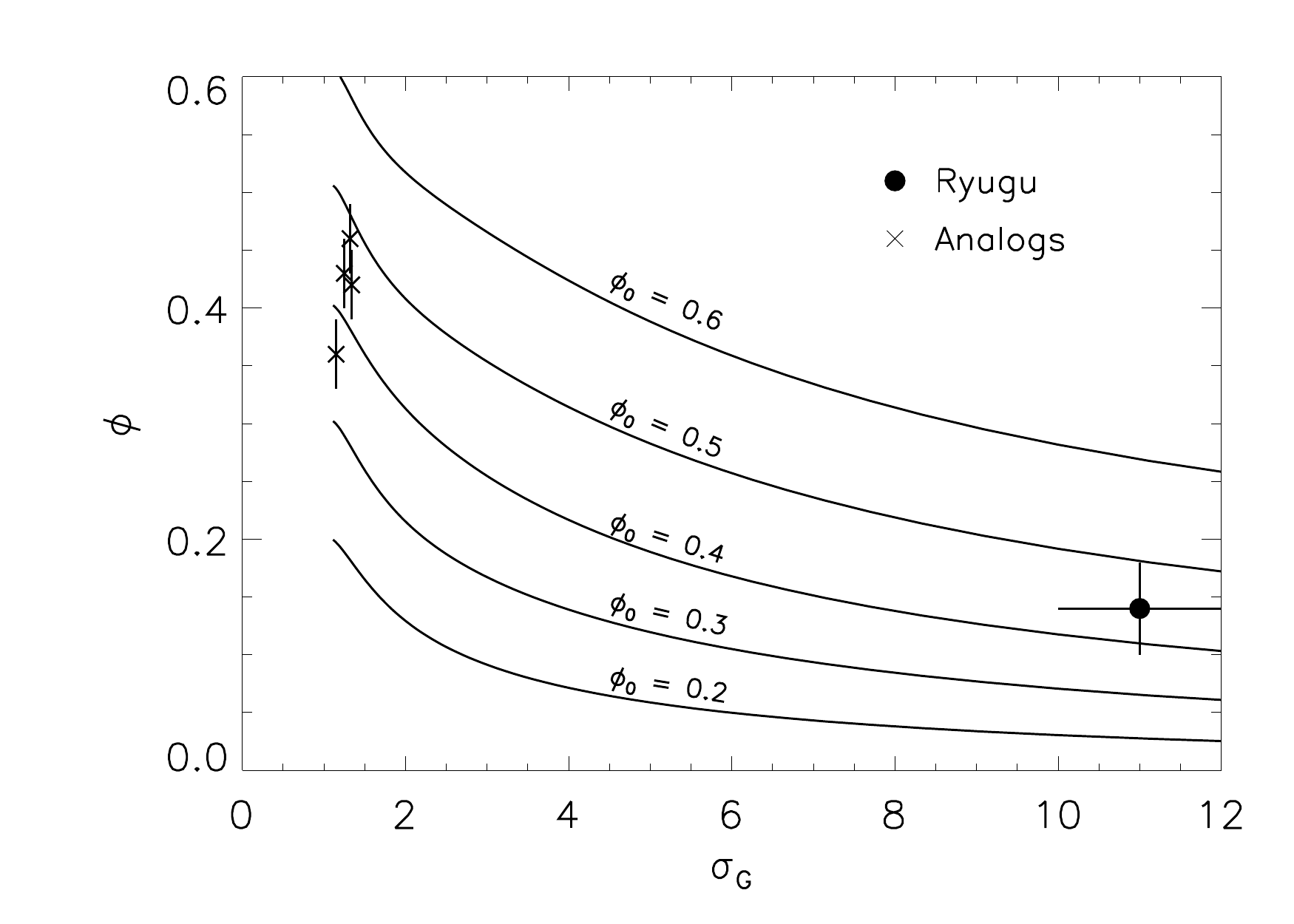}
\caption{Macroporosity ($\phi$) versus geometric standard deviation ($\sigma_G$) of a lognormal distribution for samples of varying initial porosity ($\phi_0$). The lines indicate the predictions of linear-mixture theory; this figure is an adaptation of Fig. 12 of \citet{YS91}. Data for analog samples were obtained by us (see Appendix C) and the data for Ryugu boulders, are based on the surface measurements of \citet{M19} as analyzed in this paper. } 
\end{figure}

\citet{YS91} present the porosity predictions of linear-mixture theory for a lognormal distribution (Fig. 12 of their paper) in terms of packing density (1 - $\phi$) and inverse geometric standard deviation ($\sigma_G^{-1}$). We display their results in Figure 3 in terms of $\phi$ and $\sigma_G$. Third order polynomial fits were used to transform the solution lines on their Fig. 12 to those on our Fig. 3. These lines represent the results for different values of the `initial porosity', $\phi_0$, which is the macroporosity of a pile of single-sized grains of the appropriate shape. As may be seen on Fig. 3, linear-mixture theory predicts that the macroporosity ($\phi$) of a pile decreases as the dispersion in grain sizes within it, measured by $\sigma_G$, increases. Again, this is due to the combined effects of sifting and occupation, which become more effective at eliminating pore space as the range of particle sizes increases. It is clear from the theory and its empirical confirmation that a granular aggregate with a large dispersion in grain sizes will have a smaller macroporosity than an aggregate of similarly shaped grains with a smaller range in sizes. It is also clear that to achieve a porosity in excess of about 0.25 in any object with a dispersion in grain size as large as Ryugu's, would require rather extremely shaped objects ($\phi_0 > 0.6$). 

The value of $\sigma_G$ for Ryugu is obtained from the lognormal fit to its boulder-size distribution discussed in Section 3. According to the formulation used by \citet{YS91} it is $\sigma_G = exp(\sigma) = 11.0 \pm 1.0$. Therefore, we only need to determine $\phi_0$ to predict Ryugu's macroporosity from linear-mixture theory. Initial porosity can be determined empirically from laboratory simulations using analog particles with a similar shape distribution to what is measured for the asteroid. Again, unless the grain shapes are extreme or there are non-gravitational forces at play, initial porosity is generally around 0.4, only a little larger than what is found for spheres in RCP \citep{YZ98}.  

As described in Appendix C, we assembled four sets of analog grains with small grain size dispersions and shapes similar to those measured for Ryugu's rocks by \citet{M19}. We measured their values of porosity in order to constrain the value of $\phi_0$ for the asteroid. The results are given in Table 2 of Appendix C and plotted in Fig. 3 as the x's. As expected they yield values for initial porosity between 0.36 and 0.48. On the basis of those measurements we estimate $\phi_0 = 0.43 \pm 0.07$ for the asteroid, where the error bar is a reasonably conservative outer bound, as evident on the figure. The solid circle on Figure 3 represents Ryugu based on these numbers. As can be seen on the figure, linear-mixture theory predicts a macroporosity of $\phi = 0.14 \pm 0.04$ for the asteroid. 

 There are several factors that could undermine the reliability of this result and deserve discussion before moving on. One, of course, is the assumption of homogeneity. Due to granular convection and the Brazil Nut Effect, it is possible that the surface layer of Ryugu is overpopulated with large boulders compared to its interior. Stratification of that sort would tend to increase $\phi$, by reducing the impacts of sifting and occupation. If this affects only the largest boulders, e.g. Otohime, which perches precariously on the very edge of the surface layer, it would not have much effect on $\phi$ since, as Fig. 2 shows, the largest boulders account for only a very small fraction of the total mass. In linear-mixture theory, it is the intermediate-size grains that primarily control the macroporosity of the pile and it is unclear whether granular convection and shock-induced shaking and mixing could stratify an asteroid sufficiently to have a significant effect on $\phi$. Exactly how a granular aggregate asteroid evolves, given repeated impacts that can create smaller fragments on the surface, which will sift downward, displacing larger boulders that are forced upward, is an interesting question. Computer simulations of this process have been reported by \citet{M14}. Depending on how long Ryugu has existed as a granular aggregate it may have `turned over' once or thousands of times. Rocks on its current surface have not necessarily been there since its formation and the object likely transforms itself constantly in response to impacts.
 
 Unfortunately, there is no simple way at present to test whether Ryugu is actually homogeneous or, more precisely, to determine exactly how inhomogeneous it is. In a granular aggregate there is reason to expect that it will evolve toward inhomogeneity since repeated shaking, driven by impacts, may lead to stratification, with the smaller particles sifting to the bottom. This would increase porosity both there and at the surface due to a smaller range in particle sizes. \citet{S20} has presented some weak ($\sim3 \sigma$) evidence of such an effect on Bennu, which is about half the size of Ryugu. On Itokawa, one finds a `beach" of smaller grains at the gravitational low point between its two lobes, indicating that a sifting-driven stratification effect is present on its surface. If Ryugu's interior has evolved to a more stratified state, that would increase its macroporosity over the value determined here on the basis of an assumption of homogeneity. On the other hand, repeated surface impacts might reduce the macroporosity at the surface by increasing the size dispersion through fracturing. While we see no way at present to correct our estimate for potential effects of inhomogeneity, if the density of surface rocks in the sample returned by the Hayabusa2 mission matches our prediction of $1.38 \pm 0.07$ gm cm$^{-3}$ for the average density of rocks through its entire volume, it will support the assumption of homogeneity. If the density of surface rocks in the sample return is much different from the prediction then a likely cause would be inhomogeneity of the asteroid, although it would probably require another spacecraft mission to prove that. 

Another concern relevant to the determination of $\phi$ is the mobility of grains. If cohesive forces prevented sifting one could, perhaps, end up with an asteroid having a hard outer shell and substantial voids within. On Earth, cohesive forces do not become important until d $\leq$ 0.15 mm \citep{Z11}, but in the much weaker gravity field of an asteroid, it is possible that cohesion is important for larger objects. \citet{G20} discuss this issue in depth and estimate 0.52 m as the maximum size of boulders that could be affected by cohesion on Ryugu. They describe its effects on their calculation to be of secondary importance at most. One way of testing this is to search for evidence of mobility on the surfaces of rubble pile asteroids and examples of this are growing. Bennu, for example, has been observed to eject small particles from its surface and shows other clear examples of the mobility of its surface material \citep{J20}. \citet{R19} have noted the relaxed nature of surface features on many rubble pile asteroids, which suggests there are no mobility issues, at least among the meter-scale boulders that control macroporosity on Ryugu. Cohesion effects can formally be taken into account in linear-mixture theory by adjusting the initial porosity upwards to account for them. The conservative error bars that we place on $\phi_0$ in Fig. 3 reflect an allowance for that. Absent any definitive demonstration that mobility or cohesion effects are important on asteroids at any scale, we cannot go further than that in adjusting our results to accommodate them. 

In the late stages of this study we learned of a similar effort to determine the macroporosity of Ryugu by \citet{G20}, and their results are now published. They report a value of $\phi = 0.16 \pm 0.03$, in perfect agreement with what we have found. As in this paper, they employed the linear-mixture theory of \citet{YS91}, but in its form appropriate to discrete, rather than continuous, frequency distributions. They also employed the surface boulder counts of \citet{M19}, but use a different approach to the transformation from areal to volume counts. As noted above, they also consider in detail the possible influence of cohesive forces on their results, and find them to be negligible. 

To summarize, we find the macroporosity of Ryugu, based on its surface layer grain-size distribution and an assumption of homogeneity to be $\phi = 0.14 \pm 0.04$. This corresponds to an average density of $1.38 \pm 0.07$ gm cm$^{-3}$ for its rocks (grains), throughout its volume, which is consistent with the results obtained for its surface rocks, based on their thermal properties and an independent study of the macroporosity of Ryugu \citep{G19,G20, O20}. These results indicate that the low bulk density of Ryugu is due to a very low density for its constituent rocks, not to an inefficient packing of them. Since the required rock density is much lower than any chondrule-bearing meteorite and only marginally consistent with any meteorite class at all (CI's), this raises the issue of how representative Earth-based meteorites may be as samples of primitive asteroidal material. In particular, it appears that our collections on Earth may overrepresent dense, better-lithified objects such as most chondrite classes (and the chondrules they contain) by a substantial amount. One consequence of this may be that we have overestimated the abundance of chondrule-bearing chondrites and the chondrules themselves within the Solar System, thereby exaggerating their importance to asteroid and terrestrial planet formation. In the next section we explore this line of reasoning in a bit more detail, especially as it impacts an outstanding problem in cosmochemisty -- the origin of chondrules.  

\section{The Abundance of Chondrules in Primitive Asteroidal Material}

Chondrules are well-lithified, hard spherules with measured densities, $\rho_{ch}\sim$3.0-3.4 gm cm$^{-3}$ \citep{G83,F15}, so it is clear that they cannot be the dominant component of rocks whose overall density is only 1.38 gm cm$^{-3}$. The rare CI's are the only chondrite class known with a member approaching such a low value, 1.57 gm cm$^{-3}$ \citep{M11}, and they have no chondrules. While other factors, such as chemical composition, certainly affect the bulk density of chondrites, the importance of chondrule abundance is obvious. Most classes of chondrites are quite rich in chondrules, with volume fractions, x, in the range 50 - 80\%, corresponding to mass fractions, f$_{ch}$, of 67 - 90\%. The exceptions are the CM class, with x $\sim$ 10\% (f$_{ch}$ $\sim$30\%) and the CI class with x = f$_{ch}$ = 0\% \citep{S04}.   

Fig. 4 shows the rock densities measured for several classes of CC and OC meteorites by \citet{M11} along with an indication of their chondrule content. The expected correlation between them is clear. We also display on the figure the range of possible average rock densities applicable to the three rubble pile asteroids that have been visited by spacecraft. These depend on the macroporosity of the asteroids, through Equation 2. Open bars represent the range of possibilities corresponding to $0.0 < \phi < 0.4$, while the solid portion of Ryugu's bar shows the allowed portion of its range found above, $0.10 < \phi < 0.18$. These data suggest that chondrules may be much less common on Ryugu than they are within any chondrule-bearing meteorites, including the CM class.

It is important to note that the implied low density for Ryugu surface rocks does not guarantee that they are deficient in chondrules, since their matrix may be unlike any reference material on Earth. For a two-component chondrule/matrix system, we may write that 
\begin{equation}
\rho_{rock} = f_{ch} \rho_{ch} + (1 - f_{ch}) \rho_{matrix}.
\end{equation}

\noindent Since the inferred Ryugu rock density is even less than that of CI meteorites (the lowest density chondritic material on Earth) $f_{ch} = 0$ would be required if Earth-based samples were an accurate guide. However, Ryugu rocks could have even lower matrix density than CI meteorites, if their microporosity were especially high or their grain density especially low, or both \citep{G20}. Hence, we must await the analysis of Hayabusa2 samples to provide any firmer constraint on the fractional mass abundance of chondrules in primitive asteroidal material. While we have not analyzed Bennu in detail, its overall similar appearance to Ryugu and identical bulk density suggest it is also a granular aggregate with a large value of  $\sigma_G$. If so, its macroporosity and average rock density should be similar to that of Ryugu, and one might expect a similar value of $f_{ch}$. 

\begin{figure}[ht!]
\plotone{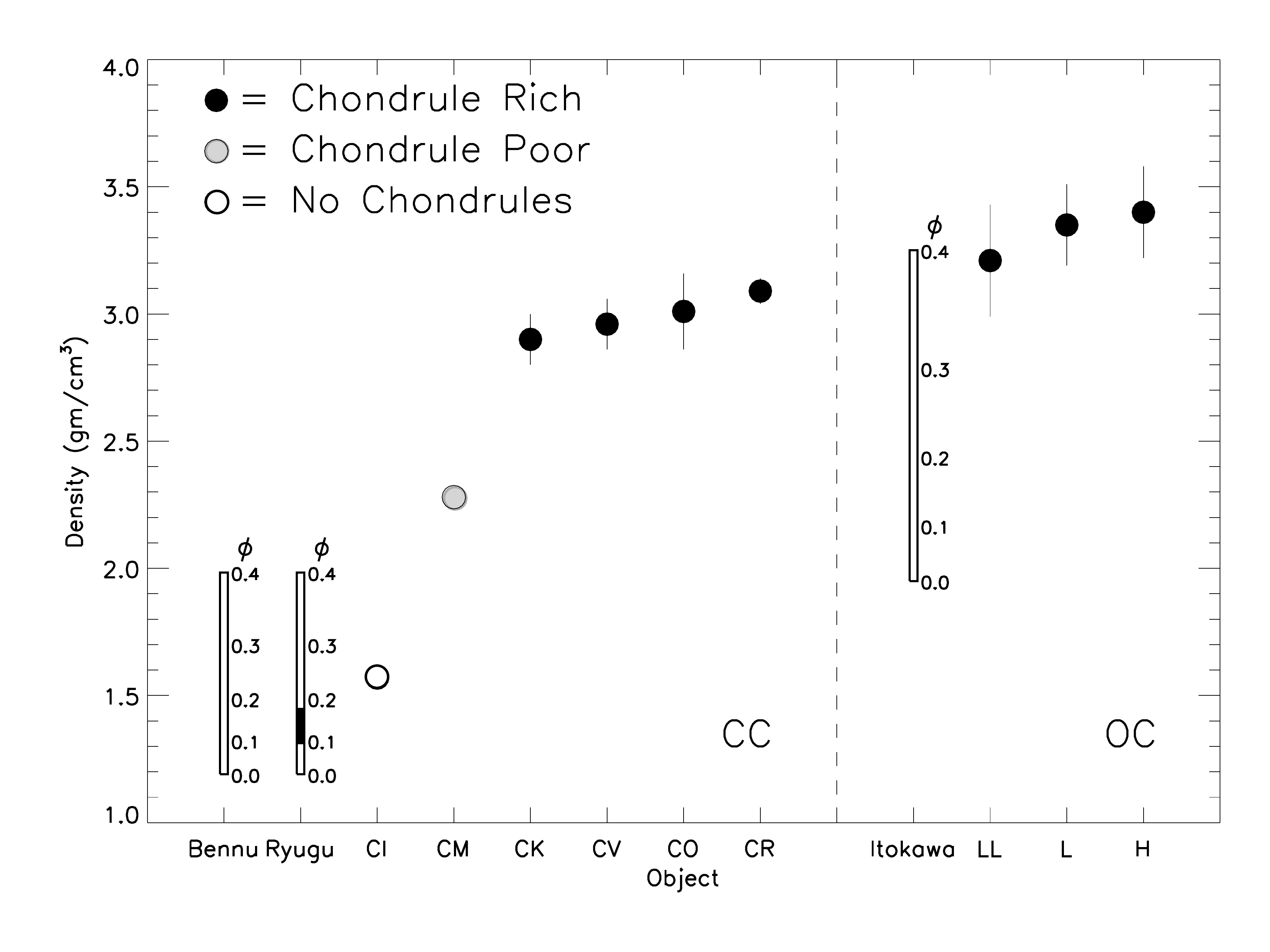}
\caption{The density of some classes of carbonaceous chondrites (CC) and ordinary chodrites (OC) from \citet{M11} are compared to the average density of rocks on three rubble pile asteroids as a function of their macroporosity ($\phi$; open bars). The result from this study that $\phi = 0.14 \pm 0.04$ is indicated by the solid portion of Ryugu's bar and corresponds to a density for its rocks of $1.38 \pm 0.07$.} 
\end{figure}

The situation with respect to the S-type asteroid Itokawa and the OC's is less clear. The power law fit to its areal boulder counts is much different than for Ryugu \citep{G20}, but its macroporosity may still be in the range 0.1-0.2, since this depends only on $\sigma_G$ and $\phi_0$ if $f(d)$ is lognormal. That would imply an average rock density of $\sim$2.3 gm cm$^{-3}$, which is significantly higher than Ryugu, but still much lower than any OC meteorite class. The higher density of OC meteorites in general (see Fig. 4) reflects not only their greater abundance of chondrules, but also a difference in their overall composition, which in turn reflects their formation in a warmer zone of the solar nebula, closer to the Sun \citep{V20}. The relative lack of volatiles in OC material and S-type asteroids may be responsible for the absence of classes of aqueously altered meteorites among the OC's that would be analogous to the CM and CI classes.  

Is the small fractional abundance of chondrules suggested by our results for Ryugu at variance with the meteorite record, which clearly shows that chondrules dominate Earth-based samples of primitive asteroidal material? Not necessarily, since as pointed out in the introduction, the flux of chondrule-bearing meteorites, measured by what makes it to the Earth's surface \citep{B96,Z06}, is much less than the flux of material at the top of the atmosphere \citep{D17}. Most of the flux of solids at 100 km above the Earth's surface is in the form of small interplanetary dust particles (IDPs). If, as has been argued, for example, by \citet{D02}, these are mostly asteroidal in origin, then there is no conflict. It would simply mean that our Earth-based sample is heavily biased towards more lithified, chondrule-bearing material, as \citet{S98} has argued. In support of this view, we note that the spectrum of surface material on Ryugu, while not exactly matching any meteorite class, does most closely resemble heated CI's and, possibly, IDPs \citep{K19}, and that \citet{G19} have found, based on an analysis of their thermal data, that most of Ryugu's boulders would be too fragile to survive passage through the Earth's atmosphere. 

To summarize, it appears that Ryugu, and likely Bennu, are composed primarily of lower density, less well-lithified and possibly less chondrule-rich material than commonly found in our meteorite collections. If most IDPs are asteroidal in origin, an issue which is unresolved because some could also be of cometary origin, then the ratio of meteorite to IDP impacts at the top of the Earth's atmosphere ($\sim10^{-3}$) could indicate the relative abundance of well-lithified to not-so-well-lithified primitive asteroidal material available in the Solar System at 1 AU today. We note that chondrules were probably even fractionally less abundant in the asteroid belt 4.6 Gyr ago since their enhanced density and hardness probably increases their odds of survival, as \citet{RS19}, for example, have discussed. It may be, therefore, that only a very minor fraction of the preserved chondritic material in the Solar System, perhaps $\sim10^{-3}$ or less, was processed into chondrules, despite their commonness in Earth-based meteorites.  

This issue will be informed soon by the contents of the samples being returned from Ryugu and Bennu. Since there is no reason, of which we are aware, that these samples should be unrepresentative of the full asteroid in terms of chondrule content, we predict on the basis of Fig. 4 that chondrules will make up significantly less than 10\% of the volume of those samples, and that chondrules could, indeed, be entirely missing from them. The OSIRIS-REx mission \citep{L17} has apparently obtained a large sample, of order a few hundred grams of material, from Bennu. If its rocks are representative of meteorites in our collections this would imply that they will have collected at least $\sim$100 grams of chondrules, i.e. millions of them. By contrast, based on the low rock densities found here, we predict the sample will contain thousands of chondrules, at most, and possibly none at all. Both asteroids, of course, should have had some higher-density, chondrule-bearing meteoroids implanted by collisions over their lifetimes, and there is evidence of such exogenous objects in the thermal maps of Ryugu \citep{O20}. 

\section{Discussion} 

If chondrules were as common in the early asteroid belt as they are in chondrites, then one requires an efficient, global mechanism for their formation, and that has been the objective of most work to date. If they are actually rare, as argued above, one cannot dismiss inefficient or local mechanisms, just on that account. The current lack of consensus on chondrule formation may result from a decades-long community preference for something that does not exist -- an efficient, global mechanism. The central question is whether the meteorites found on Earth are representative samples of the primitive material from which the asteroids formed, or are an unrepresentative population that experienced higher than normal lithification, which was critical to their survival? Here we discuss how a simple, but inefficient, local mechanism for chondrule and simultaneous chondrite formation can account for a broad range of observed phenomena that have challenged global models, including complementarity (see Section 6.2) in the composition of chondrules and matrix within chondrites,  uniformity of chondrule sizes within meteorite classes, the existence of compound chondrules and cluster chondrites, and chondrule formation in the presence of a weak magnetic field. Admittedly, only a small fraction of primitive matter could have been processed in the manner we describe, but that may be what is required of a successful theory. 

\subsection{Inefficient and Local Chondrule Formation}

 Decades of work has yielded many constraints, but so far no widely accepted mechanism for chondrule formation \citep{CJ16, R18}. A critical first step is to identify the source of the energy within the asteroid belt that was capable of elevating the temperature of pre-chondrule material to around 1600 - 1800 K for times on the order of an hour, since that is what is demanded by chondrule textures \citep{HR90, J18}. Other fundamental constraints are the ages of chondrules, which sets their formation epoch, the composition of the gaseous environment in which they formed, and the complementarity that exists in the composition of chondrules and matrix within chondrites.    

The oldest solids known to have formed within the Solar System are the CAI's. Pb-Pb and Al-Mg dating techniques yield absolute and relative ages, respectively, that are consistent with an age spread of $\leq$ 0.3 Myr, and that is generally used to define t = 0 for the Solar System \citep{CB12, K13, CB18, N18}. CAI's are likely to have condensed from the solar nebula at a time when it was still hot from accretion, a time which is only expected to have lasted for $\sim$0.1 Myr \citep{CY10}. CAI's are  highly refractory and fortunately many have survived, embedded in chondrites. \citet{G08} have shown that they formed in a reducing atmosphere at a pressure and temperature consistent with expectation for a protoplanetary disk dominated by H. 

Chondrules, on the other hand, formed in an O-rich gas that was also enhanced in Na and perhaps other volatiles \citep{A08, G08}. Unlike CAI's they exhibit a range of ages that extends to $\sim$4 Myr, which marks the end of the chondrule formation epoch. There is some debate about when it began. The most widely used and precise chronometer for the early solar system is based on the $^{26}$Al $\rightarrow$  $^{26}$Mg decay, which has a half-life of 7.17 $\times$ 10$^5$ yr. It yields ages relative to the CAI's, under the assumption that $^{26}$Al was uniformly distributed within the CAI and chondrule formation zone of the solar nebula. Over 150 chondrules from a variety of meteorites have now been dated by this technique and none yield a formation age earlier than t = 1.8 Myr, indicating that there was a significant gap in time between the formation of the CAI's and the formation of the first chondrules \citep{V09, K13, N18, P19}. This result is supported by work employing the $^{182}$Hf $\rightarrow$ $^{182}$Ta $\rightarrow$ $^{182}$W decay chain, which has a half-life of 8.9 Myr \citep{B16, KB18}.

The Pb-Pb system of chronometry \citep{B17} yields absolute ages under reasonable assumptions, but with somewhat less precision and, perhaps, more susceptibility to systematic error due to the possibility of environmental contamination \citep{P19}. While there is good agreement with the short-lived isotope results on the main points that CAI's are the oldest objects and have a much narrower age spread than chondrules, there is controversy over whether chondrule formation began at t = 0. At present, there are only 22 chondrules with Pb-Pb ages reported and they yield a mean age of 1.4 $\pm$ 1.3 Myr, consistent with results from the short-lived isotopes \citep{CB12, B17, CB18}. The fact that some chondrules yield dates consistent with t = 0, leads to the claim that chondrule formation began at that time, but given the small number of objects and the large error bars involved this is not yet convincing to all. In addition, reconciliation with the $^{26}$Al data requires that the isotope was not distributed evenly in the early solar system \citep{B17, CB18}, which has its own set of issues \citep{K18, P19}. While the matter is unresolved, a gap in time between CAI and chondrule formation of $\sim$1 Myr is not ruled out by currently available Pb-Pb ages, in our opinion.          

Accepting that chondrules formed in the asteroid belt between about 1.8 and 4 Myr in an oxidizing, not reducing, environment by a relatively inefficient process narrows down the possible energy sources. `Nebular' models, in which some event, such as shocks or lightning, heats the gas in which pre-chondrule material is suspended, are challenged by the required composition of the gas, the fact that the gaseous disk may have already dissipated by then, and the {\it ad hoc} nature of the posited events. `Planetary' models have mostly invoked collisional heating between colliding planetesimals \citep{JC18}. They are supported by the O-rich gas requirement, but challenged by many details of chondrule textures. In particular, it has not yet been demonstrated either experimentally or by computation that igneous spherules with the characteristics of most chondrules do actually form in a collisional process. And, since collisions between objects have continued over the entire 4.6 Gyr history of the Solar System, the mechanism begs the question of what put an end to chondrule formation. 

One potential source of energy for chondrule formation that we know was present in the asteroid belt between t = 1 and 4 Myr is the decay of $^{26}$Al. Given its canonical abundance at t = 0, there is more than enough energy in the decay of this isotope to fully melt (several times over) any object larger than 5-10 km in diameter \citep{L77, ss18}, assuming it forms early enough to trap that energy. \citet{SS12,ss18} have developed a chondrule formation theory that makes use of this accumulated energy. They showed, based on the models of \citet{HS06}, that 50-100 km radii planetesimals formed near t = 0 would be expected to develop fully liquid interiors encased by very thin crusts. They are the likely parent bodies of the iron meteorites, whose ages cluster near t = 0 \citep{SE06, KB17}. In the Sanders and Scott scenario, impacts onto these large, differentiated planetesimals (LDPs) fracture their thin crusts releasing molten material as `splash', which cools and solidifies into chondrules. The model neatly accounts for the gap between CAI formation and the start of chondrule formation as the time needed to form an LDP and fully melt its interior. It further identifies the end of the chondrule formation epoch near t = 4 Myr as the time by which the $^{26}$Al abundance has dropped sufficiently that LDP crusts will thicken and can no longer release molten material during an impact. The fact that one probably cannot form a large number of chondrules by this mechanism is not necessarily an argument against it, given the results of this study that chondrules may not be all that common in the Solar System, but, again, it remains to be proven that objects with chondrule textures and compositions can form from a splash into space of a molten asteroid's interior. And, the mechanism provides no clear pathway to explaining complementarity (see Section 6.2).

A variation on this model has been proposed by \citet{HG16, HG19} who suggest that it is near-infrared radiation from the ruptured thin crusts of the LDPs, not the material itself, that results in chondrule formation. They show that primitive, porous pre-chondrule material in the vicinity of an LDP with a ruptured crust could be heated sufficiently to induce chondrule formation as it flew by exposed incandescent lava. They calculated a series of heating/cooling curves that would be expected as a small primitive planetesimal (SPP) orbited a LDP under the influence of its gravity and, importantly, verified experimentally that objects with chondrule-like, porphyritic textures could be made in laboratory simulations based on those curves. This model potentially satisfies the constraint that the gaseous atmosphere be rich in O and Na and accounts for the epoch of chondrule formation in the same way as the \citet{SS12} model, The efficiency of the flyby mechanism depends on whether SPPs typically orbit LDPs many times before accreting, but if chondrules are rare, then it is probably efficient enough \citep{HG19}. 

The `flyby model' has several attributes that, arguably, make it uniquely attractive. The most important is that chondrite lithification accompanies chondrule formation, thereby satisfying the severe constraints of isotopic complementarity \citep{KB18}, compound chondrules, and cluster chondrites;  this is discussed in the next subsection. A second is that radiative heating may provide a natural explanation for the correlation between chondrule size and chemical composition, as shown by \citet{E95}, obviating the need for any size-sorting mechanism. The optical properties of pre-chondrule aggregates influence their degree of heating and these, in turn, depend on the chemical composition of the pre-cursors. Indeed, a size-class relationship is an expected signature of a radiative heating process. Finally, the clear evidence that chondrules formed in weak magnetic fields support the view that an LDP was involved, as \citet{E11} and \citet{S17} have noted. The flyby model provides a heat source other than collision for chondrule formation in a `planetary' setting.     

\subsection{The Lithification and Metamorphosis of Chondrites}
 
Lithification is the process of reducing the porosity of material by compaction and/or cementation. It requires heat and pressure and can be facilitated by liquids or gasses within the pores. In reducing porosity, a lithification event increases the density and tensile strength of the material and may also affect its overall chemical composition, for example, if volatiles are lost during heating. Heat and pressure were rare commodities in the asteroid belt 4.6 Gyr ago and it is not obvious how objects as well-lithified as chondrites could have formed \citep{CB99}. Even the central pressure of the largest asteroid, Ceres, which is $\sim$1 GPa, is insufficient for lithification by cold pressing \citep{CW02}. The results reported here support the view that, in fact, most primitive material never was lithified to the extent of chondrites. This opens the door for consideration of a mechanism that could lithify a small fraction of primitive material in space before it accretes to a larger body, namely hot isostatic pressing (HIP) \citep{HG19}. This idea is bolstered by the increasingly strong evidence that chondrite lithification and chondrule formation occurred simultaneously or nearly simultaneously. If so, chondrites may have actually formed in space prior to, or in conjunction with, their accretion to a larger body.    

The strongest evidence linking chondrule and chondrite formation comes from the phenomenon of `complementarity', which is increasingly well established. Some elements and even isotopes, that are relatively depleted in some chondrules are found to be enhanced within the matrix of the same meteorites containing those chondrules \citep{B16,HB18,KB18}. If a large fraction of primitive material is transformed into chondrules then one can imagine that the elements, or even isotopes, released from the chondrules during their formation, might be trapped within a zone of the solar nebula long enough to condense within matrix grains and ultimately find their way back to meteorites as they form on parent bodies. Clearly, the longer the time between chondrule and chondrite formation, the more difficult it would be to keep the atoms sufficiently confined in space. If, as argued here, only a tiny fraction of material is formed into chondrules, then there is the additional problem of isolating the relatively small number of released atoms and isotopes from the general ambient medium, where they would  quickly mix, erasing any detectable signature of complementarity. 

If chondrules are rare, then the phenomenon of complementarity appears to demand that the whole meteorite form simultaneously with chondrule formation, so that the escaped elements and isotopes can be immediately trapped and condensed to grains that populate the matrix. Is this conceivable? Yes, if we allow that chondrite lithification could occur immediately in space via hot isostatic pressing (HIP), as the chondrules form, rather than later on parent bodies \citep{HG19}. In their scenario, porous SPP's accreting to LDP's that are exposed to a radiation blast from extruding lava will lithify into chondrites, while forming chondrules within them and trapping enough of the atoms released in the process to account for the complementarity signatures of the whole meteorites. Support for this version of chondrite formation comes from the 
existence of compound chondrules and `cluster chondrites', i.e. chondrules whose shapes and packings indicate that they were still hot and plastic when they were lithified into their chondrite \citep{M12}. These arguments quantify the allowable time difference between chondrule and chondrite formation to be, at most, days and possibly just minutes. 

A consistent story then emerges, in which chondrules and chondrites represent the small fraction of material accreting to LDPs at just the right time to avoid being subsumed in lava oceans but still have a chance of suffering an extreme heat blast during their accretion. They must land, of course, on a permanently solid surface of the LDP or, again, they will lose their distinctive primitive characteristics. One might expect, of course, that the surface will only cool gradually with time and metamorphosis of some of the meteorites will occur, as observed in the record. Meteorites that did not lose too many volatiles during their lithification heating event, could also suffer aqueous alteration on the parent body as its surface cools, although aqueous alteration during the HIP event is also conceivable. Additional work on this seemingly promising chondrite lithification mechanism may clarify that issue.  

\subsection{The Origin of Rubble Pile Asteroids}

We finish by summarizing how one implication of our results, that chondrules  possibly are and always were, a rare form of matter in the Solar System, might fit within the general picture of asteroid formation, particularly rubble pile asteroids such as Ryugu. Dating of iron meteorites \citep{SE06, KB17} shows that asteroids large enough to melt their cores formed within a few hundred thousand years of t = 0. Simulations of asteroid growth by collisional accretion \citep{W19} indicate that most matter accretes to large bodies within a million years or less, well before the start of chondrule formation at t = 1.8 Myr, as inferred from their $^{26}$Al ages \citep{N18, P19}. In the standard picture, these large planetesimals either collide to form embryos that build terrestrial planets or are ejected from the Solar System by resonance interaction with Jupiter. According to the simulations of \citet{W19}, most of the initial mass in the asteroid belt is either already incorporated into a body large enough to melt or ejected from the Solar System before the chondrule formation epoch even begins. On the basis of their age distribution and the pace of planetesimal growth indicated by accretion models, we can conclude that chondrules (and chondrites) most likely played no significant role in building asteroids or planets.

Instead, chondrules appear to be a feature of the final sweep-up phase of planetesimal formation occurring between t = 1.8 and 4 Myr. and involving $\sim$10\% or less of the total mass remaining in the asteroid belt at that time \citep{W19}. This late-arriving material may accrete directly to the surface of LDP's or be stored indefinitely in the form of gravitationally bound moonlets. As \citet{E11} have proposed, a chondritic crust may accumulate on top of planetesimals that have already differentiated or are in the process of doing so. We propose that during this final phase of accretion, within the narrow time frame permitted by the ages of chondrules, a small fraction of the accreting material ($\leq 10^{-3}$) suffers a heat blast from hot lava exposed on the LDP's crust. This chance occurrence results in the formation of chondrules and chondrites. The evidence that many chondrules formed in the presence of a weak magnetic field supports this view, as LDP's are a natural source of such fields \citep{E11, S17, F18}. 

The scenario described here suggests potential solutions to a number of problems that have challenged cosmochemists over the decades besides just the origin of chondrules. We discussed above how the phenomenon of complementarity, the existence of cluster chondrites and the evidence for chondrule formation in the presence of weak magnetic fields fit in. There is also a well-known correlation between chondrule size and chondrite class that has proven difficult to explain in the standard model of chondrite formation since it requires a highly efficient size-sorting mechanism if chondrites actually lithified on parent body asteroids from an accumulation of material that independently accreted to them. In our model, the uniformity of chondrule size within a meteorite would simply reflect a degree of homogeneity within the parent SPP. It may also be that the optical properties of the pre-chondrule material can account for the variation in mean size between chondrule classes, as \citet{E95} have proposed. As they have argued, a correlation between chondrule size and chemical composition may actually be a distinctive sign of radiative heating. 

Finally, our model has implications for the origin of `rubble pile' asteroids such as Ryugu. Since its rocks are apparently not well lithified there is no need for Ryugu to ever have been part of a larger object, which was broken apart into `rubble' and then reassembled into a gravitationally bound pile, although it may have been \citep{N21}. Another possibility is that the asteroid may simply have assembled itself out of primordial blocks, which were never lithified to the degree of most meteorites, in which case `rubble pile' would be inappropriate terminology. Asteroids as small as Ryugu do need to be gravitationally attached to larger objects in order to avoid transport within the Solar System. But they could have spent most of their lives as moonlets around large asteroids. With sample returns on the way and other asteroid missions in various stages of development, this is an exciting time for the field and significant advances in our understanding of asteroid structure and formation may be expected soon. 
 
\section{Summary and Conclusions}
 
We have found that the boulder size distribution in the surface layer of Ryugu is well fit by a lognormal distribution (Eq. 3) with $\mu = 0.2 \pm 0.05$ and $\sigma = 2.4 \pm 0.1$, based on the counts of \citet{M19}. Assuming that the asteroid is an homogeneous granular aggregate, we employ linear-mixture packing theory \citep{YS91} to predict that its macroporosity is $\phi = 0.14 \pm 0.04$, consistent with the result of \citet{G20}. This indicates that Ryugu's constituent rocks must have a very low density, 1.38 $\pm$ 0.07 gm cm$^{-3}$, again consistent with the thermal properties of its surface boulders reported by \citet{G19} and \citet{O20}. Such low rock density is not found among meteorites on Earth that contain chondrules, and is below even the CI class, which has no chondrules. Given that chondrule densities range from 3.0 - 3.4 gm cm$^{-3}$, they may be absent or very uncommon in Ryugu material. If confirmed by analysis of the sample returns, this will support the view that chondrules and chondrites, far from being the `building blocks' of asteroids and terrestrial planets, are actually a rare form of primitive material preserved in the Solar System. They could be highly overrepresented on Earth because most primitive material does not have the tensile strength to survive passage through the Earth's atmosphere. 

Based on these findings, we favor the view that chondrules and chondrites formed during the late stages of asteroid formation, when a small subset of primitive material was subjected to intense infrared radiation from the hot surface of a planetesimal to which it was accreting. Such lithification events could transform weak, low density IDP-like material into hard, dense chondrites full of chondrules, which then accrete, fully formed, to the larger bodies, where they may experience metamorphosis. The chondrule and chondrite formation epoch is, therefore, defined by that narrow window of time, t = 1.8 - 4 Myr, when large asteroids were nearly fully molten and lava could easily extrude onto their surfaces, if ruptured.  The suggested lithification mechanism for the chondrites is hot isostatic pressing (HIP) \citep{HG19}. This model is consistent with many known properties of chondrites, including their formation epoch, chondrule textures, complementarity, cluster chondrites, remnant magnetism, and others. It would be inconsistent with a widespread occurrence of chondrules in the Solar System, if that is the ultimate result of ongoing sample return missions.

\section{Acknowledgments}

We gratefully acknowledge funding for this work from NASA research grant NNX17AE26G (PI: Greenwood) and from seed grants through the NASA CT Space Grant Consortium to both Herbst and Greenwood. We further thank the NSF for support of the SEM at Wesleyan through MRI grant 1725491 (PI: Greenwood). One of us (TEY) was supported by the Keck Northeast Astronomy Consortium (KNAC), which runs an REU program funded by NSF grant AST - 1559865 to Wesleyan University (PI: Seth Redfield). Some materials were provided for this work by a grant from Colgate University to one of us (TEY). We thank Martha Gilmore, head of the Planetary Science Group at Wesleyan, and Wesleyan students Alex Henton, Michael Henderson and Tristan Stetson, as well as Colgate students Ian Bania and Katie Chapman for their assistance. We are further indebted to Joel LaBella and Jim Zareski of the Wesleyan E\&ES department, Kenichi Abe, formerly of Wesleyan, and Paula Herbst for their support of the work. Finally, we thank Matthias Grott and Jens Biele of the German Aerospace Center and Tatsuhiro Michikami of Kindai University, for making their Ryugu boulder count data available to us in convenient form and Jens Biele for some very helpful comments on the work and manuscript. We are grateful to two anonymous referees and to the PSJ editor, Edgard G. Rivera-Valent\'in, for their comments and assistance during the refereeing process. 

\newpage

\appendix

\section{Converting boulder counts from areal to volumetric} 

The primary factor affecting $\phi$ for a granular aggregate is the size distribution of its constituent boulders \citep{YS91,F11}. In linear or linear-mixture packing models porosity is calculated in terms of the normalized volume-frequency size distribution of the particles, {\it f(d)}, where {\it d} is a characteristic size; here {\it d} is the average of the a and b dimensions of a boulder, as described below. If boulders from a representative volume sample ({\it V}) are sorted into $N_{bins}$ size-bins of widths $\Delta d$ and the volume of a single particle of size $d$ is $v(d)$, then the value of $f(d)$ for each bin is given by
\begin{equation}
f_i(d) \Delta_i d = {C n_i(d) v(d)}, 
\end{equation}
\noindent where $C$ is a unitless normalization constant and $n_i$(d) is the number of boulders per unit volume having a size that puts them in the bin. Note that $f(d)$ has units of m$^{-1}$. $C$ is chosen such that $\sum f_i(d) \Delta_i d = 1$, where the sum extends over all bins, $i = 1$ to $N_{bins}$. If the particles have a common density, independent of size, then $f(d)$ is also the mass-frequency size distribution of the particles and $f_i(d)$ is the fraction of the total mass of the sample that is in each bin. The normalization constant is:
\begin{equation}
 C =  \left( {\sum_{i = 1}^{N_{bins}} n_i (d) v(d)} \right)^{-1}.
\end{equation} 

The main obstacle to using packing theory to estimate the porosity of an asteroid is that we do not yet have a direct measurement of $f(d)$ based on a volume sample. However, for at least one asteroid (Ryugu), we do have boulder surface counts that extend over more than three orders of magnitude in size based on spacecraft images obtained at a series of different heights above its surface \citep{M19}. In Figs. 2b and 5b of their paper, the authors report the relative size-frequency distribution ($R$), by volume (or mass for constant density), of boulders within a given area (A) on its surface as a function of size ($d$). According to their definition,
\begin{equation}
R_i(d) =  {N_i(d) \left({d^3 \over \Delta_i d}\right)}
\end{equation}
\noindent where $N_i$(d) is the number of boulders per unit area in the $i^{th}$ bin, which has width $\Delta_i d$. Note that, unlike $f(d)$, $R(d)$ is unitless.

\citet{M19} also provide information on the shape of the boulders in their images, in terms of their a, b, c dimensions, where a is the longest dimension, b is the next longest dimension, measured in an orthogonal direction, and c is the shortest dimension, measured in a direction orthogonal to the ab plane. The boulder size ($d$) is defined by them as the arithmetic average of a and b, i.e. $d$ = (a + b)/2. 

\begin{figure}[ht!]
\plotone{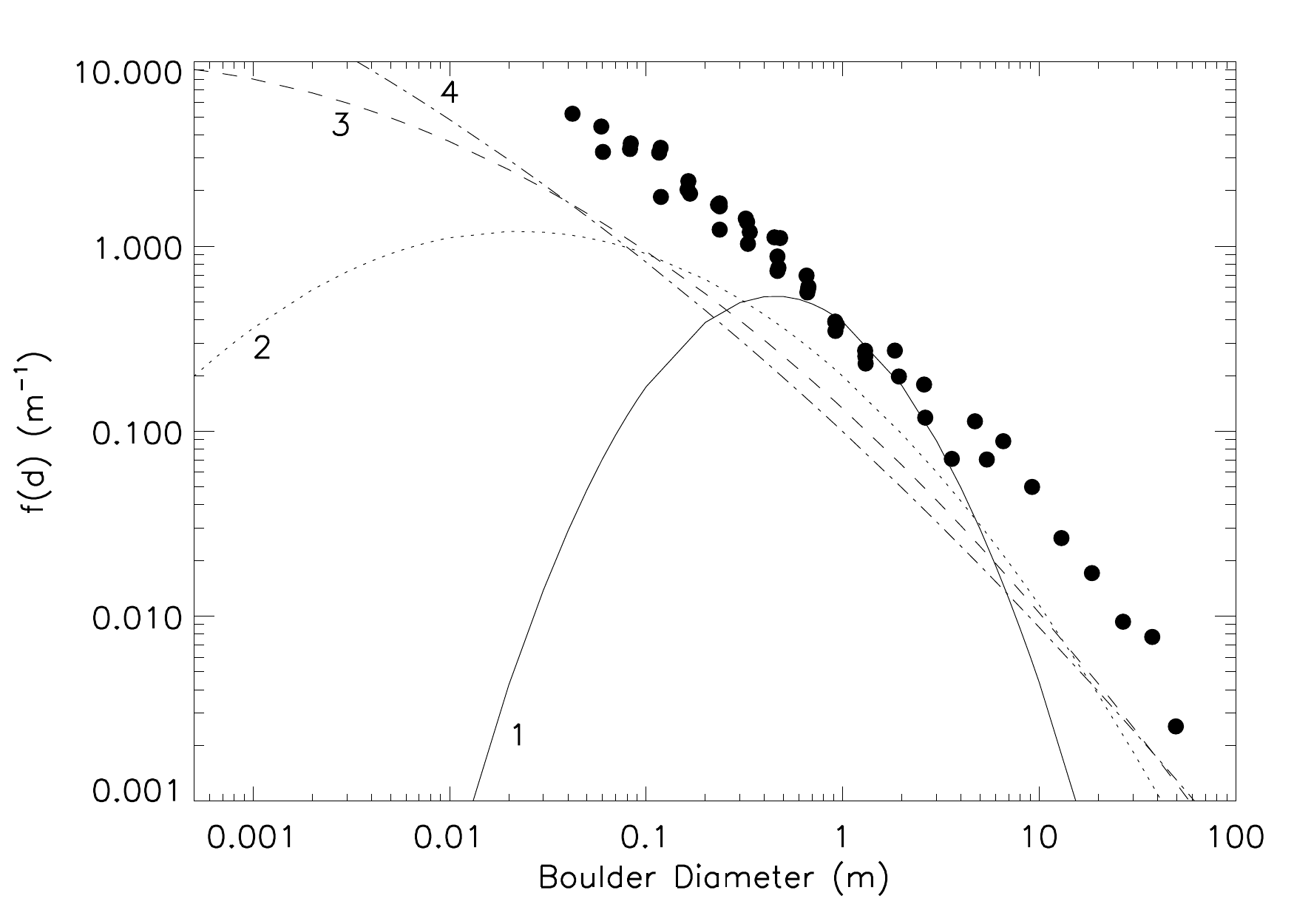}
\caption{The volume frequency distribution for Ryugu boulders, derived from the areal counts of \citet{M19}, compared to lognormal distributions with $\mu = 0.2$ and four different values of $\sigma$, as labeled. It is apparent that a value of $\sigma$ between 2 and 3 is required to fit these data.} 
\end{figure}

To derive a volume-based size distribution such as $f_i$(d), from area-based counts, represented by $R_i$(d), it is necessary to consider the thickness ($t$) of the surface layer over which boulders have been counted. Clearly,
\begin{equation}
n(d) = {N(d) \over t}
\end{equation}

\noindent but $t$ must vary with the size of the boulder, since larger boulders need not be as close to the surface as smaller boulders (e.g. pebbles) to be seen and counted. If areal counts were used directly to define the size distribution of boulders in a representative volume of the asteroid, one would over-estimate the number of large boulders. In general, $t$ can be written as $t = k \times d$, where $k$ is a unitless constant that depends on the shape and orientation of the surface boulders. As an example, if Ryugu's boulders were preferentially oriented such that their smallest dimension, c, were uniformly directed downward (i.e. $t = c$), then, based on the mean ratios of the a, b, c dimensions ($<b/a> = 0.71, <c/a> = 0.44$) $k$ would take the value 0.514. If the boulders are, instead, randomly aligned then $k = 1$. In principle, $k$ could be a function of $d$, but lacking sufficient knowledge to constrain it, here we simply assume that it is independent of boulder size, which leads to
\begin{equation}
n(d) = {(k d)^{-1} N(d)}. 
\end{equation}

Combining equations 1 through 5 and simplifying leads to the desired relationship between $f_i$(d) and $R_i$(d), namely

\begin{equation}
f_i(d) = \left({\sum_{i = 1}^{N_{bins}} R_i (d) \Delta ln (d)}\right)^{-1} \left({R_i (d) \over d}\right)
\end{equation}
Note that this equation applies independently to each of the six areas (A through F) surveyed by \citet{M19}; i.e. the sum over all bins refers to all bins within a particular area. Also note that the volume of an individual boulder depends on its shape, i.e. $v(d) = k_{shape} d^3$, so an implicit assumption in the derivation of Equation 6 is that shape is independent of size.

We calculated $f_i$(d) for Ryugu from the original boulder counts of \citet{M19}, kindly supplied by the authors. We used 8 bins of equal logarithmic width $\Delta log(d) = 0.15$ for each area and chose the lower size boundary so as to avoid the turnovers that occur in all the cumulative counts due to incompleteness at the smallest sizes. Our derived values of $f_i$ are plotted in Fig. 5. We caution that these values cannot be directly compared with each other because the normalization is specific to an area, but nonetheless the figure reveals the basic character of $f(d)$. It is a broad distribution rising to the smallest boulders counted. Intriguingly, it also follows a zero-mean lognormal distribution with a $\sigma$ of around 3 rather well, {\it albeit} with an offset due to the differing normalizations of the functions. To proceed, we must re-normalize the data, as described in the main text and Appendix B. 

\section{Renormalization and fit to a lognormal distribution}

\begin{deluxetable}{cccc}
\tablecaption{Renormalization Factors}
\tablehead{\colhead{Area\tablenotemark{a}} & \colhead{d$_{min}(m)$} & \colhead{d$_{max}(m)$} & \colhead{C$_{LN}$ }}
\startdata
A & 4.0 & 63.0 &  0.26 \\
B & 0.40 & 6.30 &  0.43 \\
C & 0.20 & 3.20 &  0.43 \\
D & 0.10 & 1.60 &  0.39 \\
E & 0.05 & 0.79 &  0.33 \\
F & 0.035 & 0.56 &  0.30 \\
\enddata
\tablenotetext{a}{as defined by \citet{M19}}
\end{deluxetable}
The similarity of $f_i$(d) to a lognormal distribution leads us to consider more quantitatively how well the data fit this uniquely important function. To make such a comparison we need to re-normalize the $f_i (d)$ values to a common system, based on the normalization of the lognormal distribution, to which  $\int_0^\infty f(d^\prime) dd^\prime = 1$ applies. We proceed by multiplying the $f_i (d)$ values within a given size range, as appropriate to the area under consideration, by the fraction of the full size range that they represent in a lognormal distribution of the appropriate mean and sigma, i.e. by the constant, C$_{LN}$, defined as
\begin{equation}
C_{LN} = \int_{d_{min}}^{d_{max}} f_{LN}(d; \mu,\sigma) dd = {{1 \over 2} \left( {{\rm erf} \left( {{\rm ln}\ d_{max} - \mu} \over {\sigma \sqrt 2} \right)} - {\rm erf} \left( {{\rm ln}\ d_{min} - \mu} \over {\sigma \sqrt 2} \right) \right) }
\end{equation}
where $f_{LN}$ is the lognormal probability density function as written in Equation 3, and erf is the usual error function,
\begin{equation}
\rm{erf} (z) \equiv {2 \over \sqrt \pi} \int_0^z e^{-t^2} dt.
\end{equation}

The re-normalized values of $f_i(d)$ were then fit to a lognormal distribution using a weighted non-linear least squares procedure based on the Levenberg-Marquardt algorithm. The data were weighted according to the number of boulders counted within a bin, i.e. by $\sigma^{-2}$ under the assumption that counting errors dominate the uncertainties.  An iterative process of fitting, followed by re-calculating the re-normalization factors, was required, but converged rapidly. Final values of the re-normalization constants are given in Table 1. The adopted fit is characterized by $\sigma = 2.4 \pm 0.1$, $\mu = 0.20 \pm 0.05$ and a reduced $\chi^2 = 1.13$. It is shown in Fig. 1. See the main text of the paper for further discussion of this result.

\section{Determining Initial Porosity}

\begin{deluxetable}{cccccccc}
\tablecaption{Shape Parameters, Density and Porosity of Ryugu Boulders and Analog Grains}
\tablehead{\colhead{Sample} & \colhead{$<a>$\tablenotemark{a}} & \colhead{$<b/a>$} & \colhead{$<c/a>$} & \colhead{$\sigma_G$\tablenotemark{b}} &
\colhead{$\rho_{rock}$ (gm cm$^{-3}$)} & \colhead{$\phi$\tablenotemark{c}} & \colhead{$\phi_0$\tablenotemark{d}}}
\startdata
1 & $27.7 \pm 3.0$ & $0.72 \pm 0.16$  & $0.47 \pm 0.17$ & $1.25 \pm 0.10$ & $2.67 \pm 0.01$ & $0.43 \pm 0.03$  & $0.45_{-0.05}^{+0.02}$  \\
2 & $7.9 \pm 2.3$ & $0.72 \pm 0.17$  & $0.42 \pm 0.20$ & $1.34 \pm 0.10$ & $2.67 \pm 0.01$ & $0.42 \pm 0.03$ & $0.43_{-0.05}^{+0.04}$  \\
3 & $10.6 \pm 3.0$ & $0.68 \pm 0.22$ & $0.36 \pm 0.14$ & $1.32 \pm 0.10$ & $2.70 \pm 0.03$ &  $0.46 \pm 0.03$ & $0.48_{-0.03}^{+0.02}$  \\
4 & $45.3 \pm 6.3$ & $0.76 \pm 0.10$  & $0.45 \pm 0.10$ & $1.15 \pm 0.10$ & $2.66 \pm 0.02$ & $0.37 \pm 0.03$ &  $0.37^{+0.03}_{-0.03}$ \\
Ryugu & $1.40 \pm 0.14$ & $0.71 \pm 0.14$  & $0.44 \pm 0.14$ & $11.0 \pm 1.1$ & $1.38 \pm 0.07$ & $0.14 \pm 0.04$  & $0.43 \pm 0.07$ \\
\enddata
\tablenotetext{a}{Ryugu data are from \citet{M19}. Units are mm for analog samples and m for Ryugu.} \tablenotetext{b}{The unites of $\sigma_G$, the geometric standard deviation, are mm for the analog samples and m for Ryugu. Errors are estimated for the samples.}
\tablenotetext{c}{The porosity, $\phi$, is measured for the samples and determined for Ryugu from $\rho_0$ and $\sigma_G$, using Fig. 3, an adaptation of Fig. 12 of \citet{YS91}. The estimated errors include an allowance for edge effects and packing efficiency.}
\tablenotetext{d}{The initial porosity, $\phi_0$, is inferred for the samples from $\phi$ and $\sigma_G$ using Fig. 3, which is an adaptation of Fig. 12 of \citet{YS91}. A mean value from the analog samples is adopted for Ryugu.}
\end{deluxetable}

\begin{figure}[ht!]
\plotone{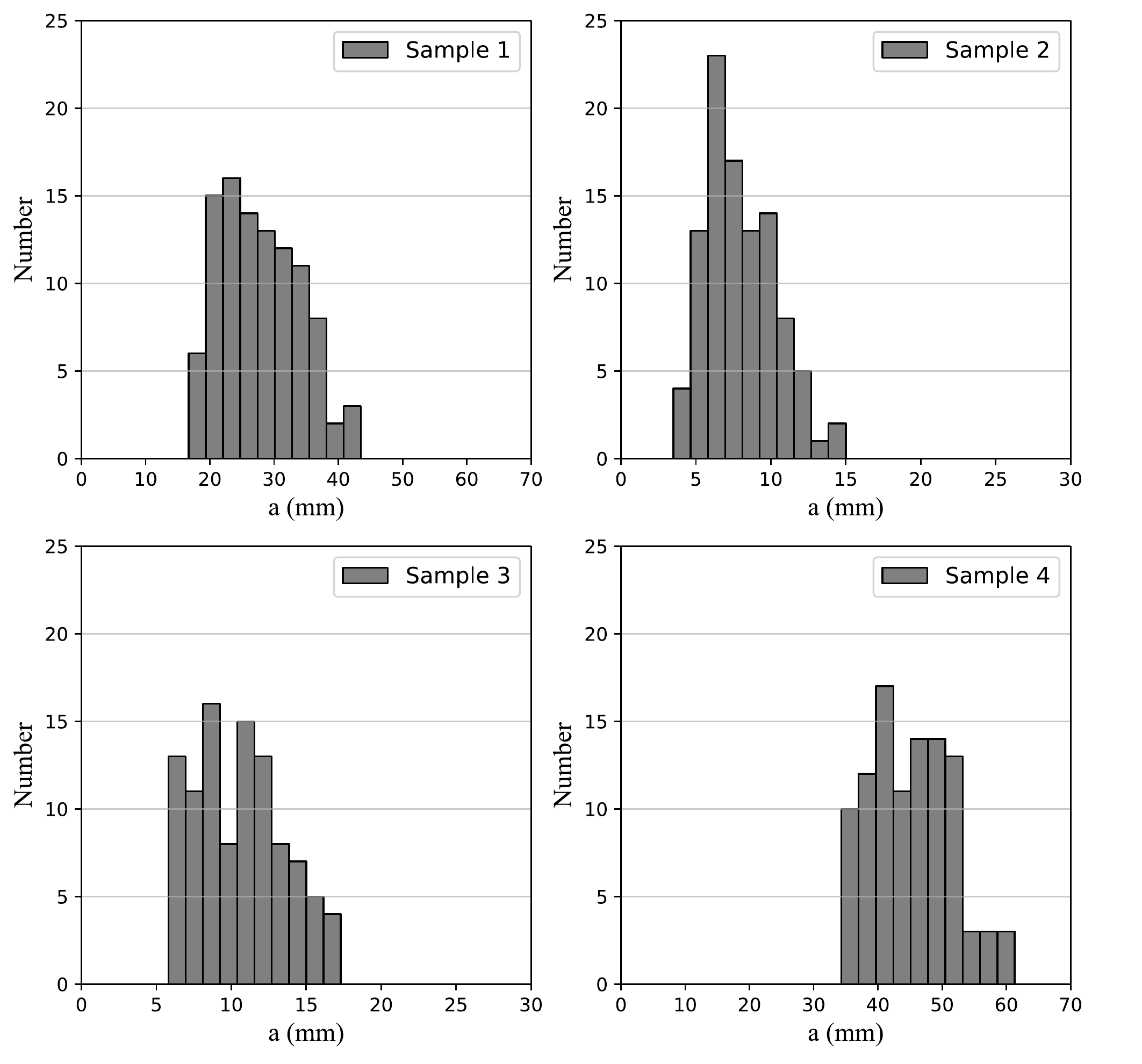}
\caption{Histograms of the `a' dimensions measured for 100 analog grains in each of our four samples.} 
\end{figure}

\begin{figure}[ht!]
\plotone{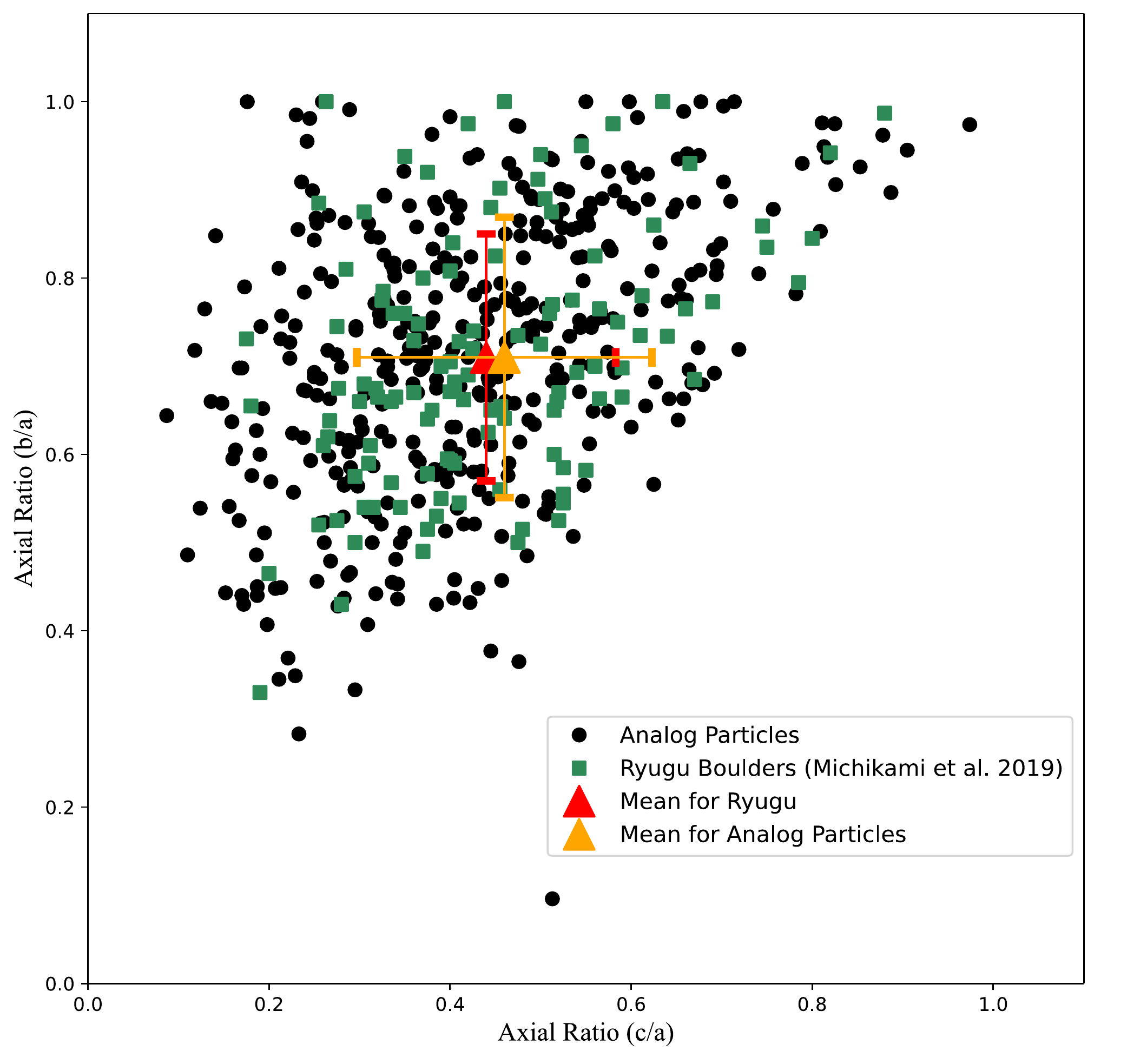}
\caption{Axial ratios for the analog particles in our four samples, as measured by us, and Ryugu, from \citet{M19}. Means and standard deviations of each group are shown. Clearly, the two sets are quite similar in shape by these measures.} 
\end{figure}

The `initial porosity', $\phi_0$, is the macroporosity of a pile of single-sized grains. It depends primarily on their shape and to a lesser degree on how they are packed, i.e. loosely or densely. One method of estimating it for Ryugu is to assemble a sample of particles with a small dispersion (measured by its geometric standard deviation, $\sigma_G$) in size that have shape parameters close to those measured for Ryugu rocks by \citet{M19}. The measured porosity for such a sample can then be transformed to $\rho_0$ using linear-mixture theory \citep{YS91}. In adopting this approach, we minimize any possible effects from cohesive forces by considering only grains that are substantially larger than 150 microns, where such forces are known to be negligible on Earth \citep{Z11}. Practical considerations of overall sample size and weight, as well as the desire to minimize edge effects, further limit us to rocks with diameters of, at most, a few cm. 

Four samples were acquired, each with a relatively small size dispersion and shape properties that approximated those reported for Ryugu boulders by \citet{M19} (see Table 2 and Figs. 6 and 7). Samples 1 and 2 are crushed rock from a quarry in Middlefield, CT, where they had been washed and size-sorted, and are sold mainly for use on road beds and driveways. Sample 3 was purchased over the Web as ``crushed gravel" and sample 4 was sold as `Mexican beach pebbles'. The beach pebbles are smoother and rounder than the crushed rock samples and yielded a somewhat smaller initial porosity than the others, perhaps on that account. No special selection or attention was made to match the measured shape parameters of Ryugu. All of our samples had axial ratios that were reasonable matches to Ryugu, regardless of size, as shown in Table 2 and the figures.

The porosity of each sample was measured in a straightforward way, by pouring the rocks into a large cylinder, shaking gently, filling to the top and weighing. From the known density of the grains, the rock volume could be determined and compared to the measured volume of the cylinder and the porosity follows from Eq. 1. By using cylinders large compared to the rocks, with heights comparable to their diameters we were able to minimize edge effects, i.e. the accumulation of pores at the rock/cylinder interfaces. The volume of the cylinders employed was 13,300 ml for samples 1 and 2 and 12,072 ml for samples 3 and 4.   

We chose a random subset of 100 rocks from each sample and measured their a, b, c dimensions to compare with Ryugu boulders. While there are small differences among our bins, the overall match to Ryugu boulders is quite good, as is evident from Fig. 7. Our analog particles have mean axial ratios of $<b/a> = 0.70 \pm 0.01$ and $<c/a> = 0.40 \pm 0.01$, which are quite close to what \citet{M19} find for Ryugu boulders, 0.71 and 0.44, respectively. As shown in Table 2, we found measured porosity to vary from 0.37 to 0.46 for our samples. Note that these are likely to be upper limits to the actual porosity of the samples because no allowance has been made for edge effects \citep{ZY96}, other than to try to minimize them in the manners described above. The measured porosities translate to initial porosites of between 0.37 and 0.48 (see Table 2 and Fig. 7). A conservative estimate for the value of the initial porosity parameter that applies to Ryugu, based on these measurements, is $\phi_0 = 0.43 \pm 0.07$, as shown in Fig. 7. That is what we have adopted in the calculations reported in the main text.


\newpage

\begin{thebibliography}{}

\bibitem[Abe et al.(2006)]{A06} Abe, S. and 15 colleagues 2006.\ Mass and Local Topography Measurements of Itokawa by Hayabusa.\ Science 312, 1344–1349.

\bibitem[Alexander et al.(2008)]{A08} Alexander, C.~M.~O. 'D ., Grossman, J.~N., Ebel, D.~S., Ciesla, F.~J.\ 2008.\ The Formation Conditions of Chondrules and Chondrites.\ Science 320, 1617. doi:10.1126/science.1156561


\bibitem[Bland et al.(1996)]{B96} Bland, P.~A., Berry, F.~J., Smith, T.~B., et al.\ 1996, \gca, 60, 2053

\bibitem[Bollard et al.(2017)]{B17} Bollard, J., Connelly, N., Whitehouse, M. J.,  Pringle, E. A., Bonal, L., Jørgensen, J. K., Nordlund, A., Moynier, F. \& Bizzarro, M. Science Advances  09 Aug 2017:
Vol. 3, no. 8, e1700407 DOI: 10.1126/sciadv.1700407

\bibitem[Britt et al.(2002)]{B02} Britt, D. T., Yeomans, D., Housen, K., \& Consolmagno, G.\ 2002, Asteroid Density, Porosity, and Structure, in Asteroids III, W. F. Bottke Jr., A. Cellino, P. Paolicchi, and R. P. Binzel (eds), University of Arizona Press, Tucson, p.485-500

\bibitem[Brown(1989)]{B89} Brown, W.K. \ 1989,  A theory of sequential fragmentation and its astronomical applications. J. Astrophys. Astr. 10, 89–112 (1989). https://doi-org.ezproxy.wesleyan.edu/10.1007/BF02714980.

\bibitem[Budde et al.(2016)]{B16} Budde, G., Kleine, T., Kruijer, T.~S., Burkhardt, C., Metzler, K.\ 2016.\ Tungsten isotopic constraints on the age and origin of chondrules.\ Proceedings of the National Academy of Science 113, 2886–2891. doi:10.1073/pnas.1524980113

\bibitem[Chiang and Youdin(2010)]{CY10} Chiang, E., Youdin, A.~N.\ 2010.\ Forming Planetesimals in Solar and Extrasolar Nebulae.\ Annual Review of Earth and Planetary Sciences 38, 493–522. doi:10.1146/annurev-earth-040809-152513


\bibitem[Ciesla(2005)]{C05} Ciesla, F.~J.\ 2005.\ Chondrule-forming Processes--An Overview.\ Chondrites and the Protoplanetary Disk 341, 811.

\bibitem[Connelly et al.(2012)]{CB12} Connelly, J.~N., Bizzarro, M., Krot, A.~N., Nordlund, {\r{A}}., Wielandt, D., Ivanova, M.~A.\ 2012.\ The Absolute Chronology and Thermal Processing of Solids in the Solar Protoplanetary Disk.\ Science 338, 651. doi:10.1126/science.1226919

\bibitem[Connelly \& Bizzarro(2018)]{CB18} Connelly, J.~N. \& Bizzarro, M. 2018, The absolute Pb-Pb isotope ages of chondrules: Insights into the dynamics of the solar protoplanetary disk, in Chondrules: records of protoplanetary disk processes. Eds: S. S. Russell, H. C. Connolly, Jr., and A. N. Krot. Cambridge University Press, Cambridge.

\bibitem[Connolly and Jones(2016)]{CJ16} Connolly, H.~C., Jones, R.~H.\ 2016.\ Chondrules: The canonical and noncanonical views.\ Journal of Geophysical Research (Planets) 121, 1885–1899. doi:10.1002/2016JE005113

\bibitem[Consolmagno and Britt(1999)]{CB99} Consolmagno, G.~J., Britt, D.~T.\ 1999.\ Turning Meteorites into Rock: Constraints on Asteroid Physical Evolution.\ Lunar and Planetary Science Conference.

\bibitem[Consolmagno, et al.(2006)]{C06} Consolmagno, G.~J., Macke, R.~J., Rochette, P., Britt, D.~T., Gattacceca, J.\ 2006.\ Density, magnetic susceptibility, and the characterization of ordinary chondrite falls and showers.\ Meteoritics and Planetary Science 41, 331–342. 

\bibitem[Consolmagno et al.(2002)]{CW02} Consolmagno, G.~J., Weidenschilling, S.~J., Britt, D.~T.\ 2002.\ Forming Well-compacted Meteorites in the Solar Nebula.\ AAS/Division for Planetary Sciences Meeting Abstracts \#34.

\bibitem[Dermott et al.(2002)]{D02} Dermott, S.~F., Kehoe, T.~J.~J., Durda, D.~D., Grogan, K., Nesvorn{\'y}, D.\ 2002.\ Recent rubble-pile origin of asteroidal solar system dust bands and asteroidal interplanetary dust particles.\ Asteroids, Comets, and Meteors: ACM 2002 500, 319–322.


\bibitem[Desch et al.(2012)]{D12} Desch, S.~J., Morris, M.~A., Connolly, H.~C., Boss, A.~P.\ 2012.\ The importance of experiments: Constraints on chondrule formation models.\ Meteoritics and Planetary Science 47, 1139–1156. doi:10.1111/j.1945-5100.2012.01357.x

\bibitem[Drolshagen et al.(2017)]{D17} Drolshagen, G., Koschny, D., Drolshagen, S., et al.\ 2017, \planss, 143, 21

\bibitem[Eisenhour and Buseck(1995)]{E95} Eisenhour, D.~D., Buseck, P.~R.\ 1995.\ Chondrule formation by radiative heating: A numerical model..\ Icarus 117, 197–211. doi:10.1006/icar.1995.1151

\bibitem[Elkins-Tanton et al.(2011)]{E11} Elkins-Tanton, L.~T., Weiss, B.~P., Zuber, M.~T.\ 2011.\ Chondrites as samples of differentiated planetesimals.\ Earth and Planetary Science Letters 305, 1–10. doi:10.1016/j.epsl.2011.03.010

\bibitem[Fuller \& Thompson(1907)]{F07} Fuller, W. B., \& Thompson, S. E. (1907). The laws of proportioning concrete. Trans. Am. Soc. Civ. Eng., 59 (2), 67-143.

\bibitem[Friedrich et al.(2015)]{F15} Friedrich, J.~M. and 7 colleagues 2015.\ Chondrule size and related physical properties: A compilation and evaluation of current data across all meteorite groups.\ Chemie der Erde / Geochemistry 75, 419–443. doi:10.1016/j.chemer.2014.08.003

\bibitem[Frings, Sch\"uttrumpf \& Vollmer(2011)]{F11} Frings, R. M., Sch\"uttrumpf, H. \& Vollmer, S. \ 2011, Verification of Porosity Predictors for Fluvial Sand-Gravel Deposits, Water Resources Research 47, https://doi.org/10.1029/2010WR009690

\bibitem[Fu et al.(2018)]{F18} Fu, R. R., Weiss, B. P., Schrader, D. L. \& Johnson, B. C. 2018, Records of magnetic fields in the chondrule formation environment, in Chondrules: records of protoplanetary disk processes. Eds: S. S. Russell, H. C. Connolly, Jr., and A. N. Krot. Cambridge University Press, Cambridge.

\bibitem[Gooding(1983)]{G83} Gooding, J.~L.\ 1983.\ Survey of chondrule average properties in H-, L-, and LL-group chondrites - Are chondrules the same in all unequilibrated ordinary chondrites?.\ Chondrules and their Origins, 61–87.

\bibitem[Greenwood, Burbine \& Franchi(2020)]{GB20} Greenwood, R. C., Burbine, T. H. \& Franchi, I. A.  \ 2020, Linking asteroids and meteorites to the primordial planetesimal population, Geochimica et Cosmochimica Acta 277, 15 May 2020, Pages 377-406. https://doi.org/10.1016/j.gca.2020.02.004

\bibitem[Grossman et al.(2008)]{G08} Grossman, L., Beckett, J. R., Fedkin, A. V., Simon, S. B., \& Ciesla, F. J. \ 2008 Redox Conditions in the Solar Nebula: Observational, Experimental and Theoretical Constraints, Reviews in Mineralogy \& Geochemistry
Vol. 68, pp. 93-140

\bibitem[Grott et al.(2019)]{G19} Grott, M. et al. \ 2019, Low thermal conductivity boulder with high porosity identified on C-type asteroid (162173) Ryugu. Nat Astron 3, 971–976 (2019). https://doi.org/10.1038/s41550-019-0832-x


\bibitem[Grott et al.(2020)]{G20} Grott, M. and 9 colleagues 2020.\ Macroporosity and Grain Density of Rubble Pile Asteroid (162173) Ryugu.\ Journal of Geophysical Research (Planets) 125. doi:10.1029/2020JE006519


\bibitem[Hales(1998)]{H98} Hales, T.~C.\ 1998.\ An overview of the Kepler conjecture.\ arXiv Mathematics e-prints.

\bibitem[Hamm et al.(2020)]{H20} Hamm, M., Pelivan, I., Grott, M., de Wiljes, J.\ 2020.\ Thermophysical modelling and parameter estimation of small Solar system bodies via data assimilation.\ Monthly Notices of the Royal Astronomical Society 496, 2776–2785. doi:10.1093/mnras/staa1755

\bibitem[Herbst and Greenwood(2016)]{HG16} Herbst, W., Greenwood, J.~P.\ 2016.\ A new mechanism for chondrule formation: Radiative heating by hot planetesimals.\ Icarus 267, 364–367. doi:10.1016/j.icarus.2015.11.026

\bibitem[Herbst and Greenwood(2019)]{HG19} Herbst, W., Greenwood, J.~P.\ 2019.\ A radiative heating model for chondrule and chondrite formation.\ Icarus 329, 166–181. doi:10.1016/j.icarus.2019.03.039

\bibitem[Hevey and Sanders(2006)]{HS06} Hevey, P.~J., Sanders, I.~S.\ 2006.\ A model for planetesimal meltdown by $^{26}$Al and its implications for meteorite parent bodies.\ Meteoritics and Planetary Science 41, 95–106. doi:10.1111/j.1945-5100.2006.tb00195.x

\bibitem[Hewins and Radomsky(1990)]{HR90} Hewins, R.~H., Radomsky, P.~M.\ 1990.\ Temperature Conditions for Chondrule Formation.\ Meteoritics 25, 309. doi:10.1111/j.1945-5100.1990.tb00715.x

\bibitem[Hezel et al.(2018)]{HB18} Hezel, D. C., Bland, P. A., Palme, H., Jacquet, E. \& Bigolski, J. 2018 Compositions of chondrules and matrix and their complementary relationship in chondrites, in Chondrules: records of protoplanetary disk processes. Eds: S. S. Russell, H. C. Connolly, Jr., and A. N. Krot. Cambridge University Press, Cambridge.

\bibitem[Jaeger \& Nagel(1992)]{JN92} Jaeger, H. M. \& Nagel, Sidney R. \ 1992, Physics of the Granular State, {\it Science} {\bf 255}, 1523-1531

\bibitem[Jawin et al.(2020)]{J20} Jawin, E.~R. and 20 colleagues 2020.\ Global Patterns of Recent Mass Movement on Asteroid (101955) Bennu.\ Journal of Geophysical Research (Planets) 125. doi:10.1029/2020JE006475

\bibitem[Johansen et al.(2015)]{J15} A. Johansen, M.-M. Mac Low, P. Lacerda, M. Bizzarro, Growth of asteroids, planetary embryos, and Kuiper belt objects by chondrule accretion. Sci. Adv. 1, 1500109 (2015).

\bibitem[Johnson et al.(2018)]{JC18} Johnson, B. C., Ciesla, F. J., Dullemond, C. P. \& Melosh, H. J., 2018 Formation of chondrules by planetesimal collisions, in Chondrules: records of protoplanetary disk processes. Eds: S. S. Russell, H. C. Connolly, Jr., and A. N. Krot. Cambridge University Press, Cambridge.

\bibitem[Jones et al.(2018)]{J18} , Jones, R. H., Villaneuve, J. \& Libourel, G. 2018 
Thermal histories of chondrules: petrologic observations and experimental constraints, in Chondrules: records of protoplanetary disk processes. Eds: S. S. Russell, H. C. Connolly, Jr., and A. N. Krot. Cambridge University Press, Cambridge.

\bibitem[Kita et al.(2013)]{K13} Kita, N.~T. and 8 colleagues 2013.\ $^{26}$Al-$^{26}$Mg isotope systematics of the first solids in the early solar system.\ Meteoritics and Planetary Science 48, 1383–1400. doi:10.1111/maps.12141

\bibitem[Kitazato et al.(2019)]{K19} Kitazato, K. and 65 colleagues 2019.\ The surface composition of asteroid 162173 Ryugu from Hayabusa2 near-infrared spectroscopy.\ Science 364, 272–275. doi:10.1126/science.aav7432

\bibitem[Kleine et al.(2018)]{KB18} Kleine, T., Budde, G., Hellmann, J. L., Kruijer, T. S., and Burkhardt, C. 2018, Tungsten isotopes and the origin of chondrules and chondrites, in Chondrules: records of protoplanetary disk processes. Eds: S. S. Russell, H. C. Connolly, Jr., and A. N. Krot. Cambridge University Press, Cambridge. 

\bibitem[Krot et al.(2018)]{K18} Krot, A. N., Nagashima, K. Libourel, G. \& Miller, K. E., 2018, Multiple mechanisms of transient heating events in the protoplanetary disk, in Chondrules: records of protoplanetary disk processes. Eds: S. S. Russell, H. C. Connolly, Jr., and A. N. Krot. Cambridge University Press, Cambridge. 

\bibitem[Kruijer et al.(2017)]{KB17} Kruijer, T.~S., Burkhardt, C., Budde, G., Kleine, T.\ 2017.\ Age of Jupiter inferred from the distinct genetics and formation times of meteorites.\ Proceedings of the National Academy of Science 114, 6712–6716. doi:10.1073/pnas.1704461114

\bibitem[Lauretta et al.(2017)]{L17} Lauretta, D.~S. and 47 colleagues 2017.\ OSIRIS-REx: Sample Return from Asteroid (101955) Bennu.\ Space Science Reviews 212, 925–984. doi:10.1007/s11214-017-0405-1

\bibitem[Lauretta et al.(2019)]{L19} Lauretta, D. S. \ 2019, The Unexpected Surface of Asteroid (101955) Bennu, Nature. 2019 Apr; 568(7750): 55–60. 

\bibitem[Lee et al.(1977)]{L77} Lee, T., Papanastassiou, D.~A., Wasserburg, G.~J.\ 1977.\ Aluminum-26 in the early solar system: fossil or fuel?.\ The Astrophysical Journal 211, L107–L110. doi:10.1086/182351


\bibitem[Limpert, Stahel \& Abbt(2001)]{L01} Eckhard Limpert,  Werner A. Stahel,  Markus Abbt  \ 2001, Log-normal Distributions across the Sciences: Keys and Clues: On the charms of statistics, and how mechanical models resembling gambling machines offer a link to a handy way to characterize log-normal distributions, which can provide deeper insight into variability and probability—normal or log-normal: That is the question, BioScience, Volume 51, Issue 5, May 2001, Pages 341–352, https://doi.org/10.1641/0006-3568(2001)051[0341:LNDATS]2.0.CO;2. 

\bibitem[Macke, Consolmagno \& Britt(2011)]{M11} Macke, R. J., Consolmagno, G. J. \& Britt, D. T. 2011.\ Density, porosity, and magnetic susceptibility of carbonaceous chondrites\ Meteoritics \& Planetary Science 46, Nr 12, 1842–1862 (2011)
doi: 10.1111/j.1945-5100.2011.01298.x

\bibitem[Matsumura et al.(2014)]{M14} Matsumura, S., Richardson, D.~C., Michel, P., Schwartz, S.~R., Ballouz, R.-L.\ 2014.\ The Brazil-nut effect and its application to asteroids.\ Asteroids, Comets, Meteors 2014.

\bibitem[Metzler(2012)]{M12} Metzler, K.\ 2012.\ Ultrarapid chondrite formation by hot chondrule accretion? Evidence from unequilibrated ordinary chondrites.\ Meteoritics and Planetary Science 47, 2193–2217. doi:10.1111/maps.12009

\bibitem[Michikami et al.(2019)]{M19} Michikami, T. and 35 colleagues 2019.\ Boulder size and shape distributions on asteroid Ryugu.\ Icarus 331, 179–191.

\bibitem[Miller and Scalo(1979)]{M79} Miller, G.~E., Scalo, J.~M.\ 1979.\ The Initial Mass Function and Stellar Birthrate in the Solar Neighborhood.\ The Astrophysical Journal Supplement Series 41, 513-547.

\bibitem[Nagashima, Kita \& Lu(2018)]{N18} Nagashima, K., Kita, N. T. \& Lu, T-H. $^{26}Al-^{26}Mg$ systematics of chondrules. in Chondrules: records of protoplanetary disk processes. Eds: S. S. Russell, H. C. Connolly, Jr., and A. N. Krot. Cambridge University Press, Cambridge.

\bibitem[Neumann et al.(2021)]{N21} Neumann, W. and 6 colleagues 2021.\ Microporosity and parent body of the rubble-pile NEA (162173) Ryugu.\ Icarus 358. doi:10.1016/j.icarus.2020.114166


\bibitem[Okada et al.(2020)]{O20} Okada, T., Fukuhara, T., Tanaka, S. et al. Highly porous nature of a primitive asteroid revealed by thermal imaging. Nature 579, 518–522 (2020). https://doi.org/10.1038/s41586-020-2102-6

\bibitem[Pape et al.(2019)]{P19} Pape, J., Mezger, K., Bouvier, A.-S., Baumgartner, L.~P.\ 2019.\ Time and duration of chondrule formation: Constraints from $^{26}$Al-$^{26}$Mg ages of individual chondrules.\ Geochimica et Cosmochimica Acta 244, 416–436.

\bibitem[Polishook and Aharonson(2020)]{P20} Polishook, D., Aharonson, O.\ 2020.\ Surface slopes of asteroid pairs as indicators of mechanical properties and cohesion.\ Icarus 336. doi:10.1016/j.icarus.2019.113415

\bibitem[Richardson et al.(2019)] {R19} Richardson, J.~E., Graves, K.~J., Harris, A.~W., Bowling, T.~J.\ 2019.\ Small body shapes and spins reveal a prevailing state of maximum topographic stability.\ Icarus 329, 207–221. doi:10.1016/j.icarus.2019.03.027

\bibitem[Rivkin and Stickle(2019)]{RS19} Rivkin, A., Stickle, A.\ 2019.\ Are There Large, Never-Lithified Asteroids?.\ EPSC-DPS Joint Meeting 2019.


\bibitem[Rosato et al.(1987)]{R87} Rosato, A., Strandburg, K. J., Prinz, F. \& Swendsen, R. H. \ 1987, Why the Brazil Nuts Are on Top: Size Segregation of Particulate Matter by Shaking, Phys. Rev. Let. 58, 1038-1040.

\bibitem[Russell, Connolly \& Krot(2018)]{R18} Russell, S. S., Connolly, H. C., \& Krot, A. N., editors 2018, Chondrules: Records of Protoplanetary Disk Processes, Cambridge university Press, ISBN: 9781108418010

\bibitem[Sanders and Scott(2012)]{SS12} Sanders, I.~S., Scott, E.~R.~D.\ 2012.\ The origin of chondrules and chondrites: Debris from low-velocity impacts between molten planetesimals?.\ Meteoritics and Planetary Science 47, 2170–2192. doi:10.1111/maps.12002

\bibitem[Sanders and Scott(2018)]{ss18} Sanders, I.~S., Scott, E.~R.~D.\ 2018.\ Making chondrules by splashing molten planetesimals: The dirty impact plume model, in Chondrules: records of protoplanetary disk processes. Eds: S. S. Russell, H. C. Connolly, Jr., and A. N. Krot. Cambridge University Press, Cambridge. 

\bibitem[Scherst{\'e}n et al.(2006)]{SE06} Scherst{\'e}n, A., Elliott, T., Hawkesworth, C., Russell, S., Masarik, J.\ 2006.\ Hf W evidence for rapid differentiation of iron meteorite parent bodies.\ Earth and Planetary Science Letters 241, 530–542. doi:10.1016/j.epsl.2005.11.025


\bibitem[Sears(1998)]{S98} Sears, D.~W.~G.\ 1998.\ The Case for Rarity of Chondrules and Calcium-Aluminum-rich Inclusions in the Early Solar System and Some Implications for Astrophysical Models.\ The Astrophysical Journal 498, 773–778. doi:10.1086/305589

\bibitem[Sears(2004)]{S04} Sears, D.~W.~G.\ 2004.\ The origin of chondrules and chondrites.\ The origin of chondrules and chondrites, by D.W.G. Sears. Cambridge planetary science series. Cambridge, UK: Cambridge University Press, 2004.

\bibitem[Sears(2017)]{Se17} Sears, D.~W.~G.\ 2017.\ Itokawa is a Regolith Breccia, Not a Rubble Pile.\ Lunar and Planetary Science Conference.

\bibitem[Scheeres(2020)]{S20} Scheeres, D. J. et al. \ 2020, Sci. Adv. 2020; 6 : eabc3350 8 October 2020

\bibitem[Scott \& Kilgour(1969)]{SK69} Scott, G. D. \& Kilgour, D. M.\ 1969, The density of random close packing of spheres, J. Phys. D. Appl. Phys. 2, 863

\bibitem[Shah et al.(2017)]{S17} Shah, J., Bates, H.~C., Muxworthy, A.~R., Hezel, D.~C., Russell, S.~S., Genge, M.~J.\ 2017.\ Long-lived magnetism on chondrite parent bodies.\ Earth and Planetary Science Letters 475, 106–118. doi:10.1016/j.epsl.2017.07.035


\bibitem[Sharma(2013)]{S13} Sharma, I.\ 2013.\ Structural stability of rubble-pile asteroids.\ Icarus 223, 367–382. doi:10.1016/j.icarus.2012.11.005

\bibitem[Sharma(2014)]{S14} Sharma, I.\ 2014.\ Stability of rubble-pile satellites.\ Icarus 229, 278–294. doi:10.1016/j.icarus.2013.09.023

\bibitem[Villeneuve, Chaussidon, \& Libourel(2009)]{V09} Villeneuve, J., Chaussidon, M. \& Libourel, G. \ 2009, Science 325, 985

\bibitem[Vollstaedt et al.(2020)]{V20} Vollstaedt, H., Mezger, K., Alibert, Y.\ 2020.\ Carbonaceous Chondrites and the Condensation of Elements from the Solar Nebula.\ The Astrophysical Journal 897. doi:10.3847/1538-4357/ab97b4


\bibitem[Watanabe et al.(2019)]{Wa19} Watanabe, S. et al. \ 2019, Hayabusa2 arrives at the carbonaceous asteroid 162173 Ryugu: A spinning topshaped rubble pile, Science 364 (6437), 268-272. 

\bibitem[Westman \& Hugill(1930)]{WH30} Westman, A.~E.~R. \& Hugill, H. R. \ 1930, The Packing of Particles, {\it Journal of the American Ceramic Society}, {\bf 13 (10)}, 767-779

\bibitem[Weidenschilling(2019)]{W19} Weidenschilling \ 2019, Accretion of the asteroids: Implications for their thermal evolution, {\it Meteoritics \& Planetary Science}, {\bf 54 (5)}, 1115-1132

\bibitem[Yang, Zou \& Yu(2000)]{YZY00} R.Y. Yang, R.P. Zou, \& A.B. Yu \ 2000, 
Computer simulation of the packing of fine particles, 
Physical Review E, 62, 3900-3908

\bibitem[Yu \& Standish(1991)]{YS91} Yu, A. B. \& Standish, N. \ 1991, Estimation of Porosity of Particle Mixtures by a Linear-Mixture Packing Model, Industrial Engineering Chem. Res. 30, 1372-1385.

\bibitem[Yu \& Zou(1998)]{YZ98}Yu, A. B., \& Zou, R. P. (1998). Prediction of the porosity of particle mixtures. KONA Powder and Particle Journal , 16 , 68-81. doi: 10.14356/kona.1998010

\bibitem[Zou et al.(2011)] {Z11} Zou, R. 718 P., Gan, M. L., \& Yu, A. B.\ 2011. Prediction of the porosity of multi component mixtures of cohesive and non-cohesive particles. Chemical Engineering Science 66, 4711-4721. https://doi.org/10.1016/j.ces.2011.06.037

\bibitem[Zou \& Yu(1996)] {ZY96} Zou, R.\& Yu, A. B.\ 1996. Evaluation of the packing characteristics of mono-sized spherical particles. Powder Tehnology 88, 71-79. https://doi.org/10.1016/0032-5910(96)03106-3

\bibitem[Zolensky et al.(2006)]{Z06} Zolensky, M., Bland, P., Brown, P., et al.\ 2006, Meteorites and the Early Solar System II, 869




\end{thebibliography}
\end{document}